\documentclass[prb,aps,epsf,twocolumn,showpacs,10pt]{revtex4-1}
\usepackage{graphicx,amsfonts,times,bm,amsmath,verbatim,color,array}

\usepackage{subfigure}
\usepackage{float}
\usepackage{epsfig}
\usepackage{psfrag}
\usepackage{hyperref}
\hypersetup{
colorlinks=true,final=true,
        linkcolor=blue,
        citecolor=blue,
        filecolor=blue,
        urlcolor=blue,
}
\begin{document}
\title{Tunable topological Nernst effect in 2D transition metal dichalcogenides}

\author{Girish Sharma}

\affiliation{Department of Physics, Virginia Tech, Blacksburg, VA 24061, U.S.A}

\begin{abstract}
Two dimensional semiconducting transition metal dichalcogenides (TMDs) exhibit an intrinsic Ising spin orbit coupling (SOC) along with a valley contrasting Berry curvature, which can generate a purely anomalous spin and valley Nernst signal driven by a thermal gradient. We show that a small Bychkov-Rashba coupling, which is present in gated TMDs, can enhance the valley Nernst signal by at least 1-2 orders of magnitude. We find that the Nernst signal in these materials is dominated by the anomalous geometrical contribution, and the conventional contribution is much weaker. Importantly, the Nernst signal is also highly tunable by external gating. Although the total Nernst signal vanishes due to time reversal (TR) symmetry,  a small magnetic coupling lifts the valley degeneracy and generates an amplified Nernst response. Additionally, we also discuss the Nernst response of bilayer TMDs, and show a similar enhancement and modulation of the Nernst signal due to Rashba SOC.
Our predictions are highly pertinent to ongoing experimental studies in TMDs. The generated large anomalous Nernst signal can directly probe the presence of a large Berry curvature in these materials, and may serve as a promising tunable platform for caloritronics applications.

\end{abstract}

\maketitle
\section{Introduction}
Recently two dimensional semiconducting transition metal dichalcogenides (TMDs) of the form MX$_2$, where M = Mo, W, and X = S, Se, have attracted enormous interest because of the possibility of manipulating the valley degree of freedom in addition to spin and charge~\cite{Xiao:2012,Xu:2014,Hao:2016,Mak:2014,Mak:2016,Yang:2015,Kormanyos:2014,Mak2010,Splendiani,Cao:2012}, making them highly promising for low-power electronics and several valleytronics applications. The valley degrees of freedom refer to local extrema in the band structures of these materials, a feature unique to the class of these materials. The valley degree of freedom may also be manipulated as a qubit which opens up new arenas to explore quantum computation with TMDs~\cite{Kormanyos:2014}.
Further, the in-plane mirror symmetry is broken and results in an unusual intrinsic spin-orbit coupling (SOC), called the Ising SOC~\cite{Cappelluti:2013,Zahid:2013,Zhu:2011,Kormanyos:2013}, which originates from the $d$-orbitals of the transition metal. Ising SOC aligns the electron spins near opposite valleys to opposite out of plane directions, thus behaving like a valley contrasting out of plane Zeeman field. Such a spin-orbit field can be sharply contrasted to the Rashba SOC~\cite{Rashba:1959} which aligns the electron spins along the in-plane direction and produces a helical liquid. As a consequence of Ising SOC the valley bands in TMDs are spin-locked even though overall time-reversal symmetry is preserved. This distinctive feature has spurred theoretical and experimental studies in semiconducting TMDs as well as their superconducting counterparts~\cite{Jonker2,Jonker3,Jonker4,Cappelluti:2013,Hao:2016,Kormanyos:2013,Lee:2016,Mak:2014,Mak:2016,Wu:2013,XXu3,XXu4,Xiao:2012,Xu:2014,Yang:2015,Zahid:2013,Zhu:2011, Sharma1,Sharma2,Splendiani,Cao:2012,Kormanyos:2014,Mak2010,XXu1,XXu2,Jonker5}. Moreover the broken spatial inversion symmetry also generates a large Berry curvature near the valleys giving rise to unconventional electronic and optical properties~\cite{Niu:2006}, which have been recently studied in the context of TMDs~\cite{Xiao:2012,Yu:2015,Zhou:2017,Mak:2014}.

In TMDs, due to spin-valley locking the electron spin remains a good quantum number. However, spin may not be conserved if the out of plane mirror symmetry is broken, which can result from a number of processes such as external gating or strain, that are typically present in realistic experimental situations. A Bychkov-Rashba spin orbit coupling~\cite{Bychkov} can thereby naturally arise in TMDs~\cite{Ochoa:2013,Kormanyos:2014} which aligns the electron spins away from the out-of plane direction. The non-trivial role of Rashba SOC has been extensively studied in several materials such as 2D semiconductors~\cite{Bychkov}, quantum dots~\cite{Tsitsishvili:2004}, more recently in  producing topological superconductivity~\cite{Lutchyn}. Despite the fact that Rashba SOC exists in gated TMDs, its role in TMDs has received only limited attention~\cite{Yuan:2014,Kli:2013,Yao:2017,Taguchi:2017}. Very recently the role of Rashba SOC in generating strong Berry curvature effects and anomalous Hall effect has been studied~\cite{Zhou:2017}. 

The Nernst effect has been used as a very important experimental probe in a number of physical systems such as high-temperature cuprate superconductors~\cite{AZXu:2000}, charge density waves~\cite{Bel:2003}, and more recently as a probe of strong Berry curvature in Weyl and Dirac semimetals~\cite{Liang:2017}. In conventional Fermi liquids, the Nernst effect is expected to be small due to Sondheimer's cancellation~\cite{Sondheimer}. However, the Sondheimer's cancellation breakdowns for Berry mediated Nernst effect~\cite{Sharma3}, giving rise to a measurable anomalous Nernst current even in the absence of magnetic field. The generation of a Nernst signal in TMDs has been investigated where the intrinsic Ising SOC and Berry curvature effects can generate a pure spin and valley Nernst current~\cite{Yu:2015}. 

In this work we show that extrinsic Rashba SOC and intrinsic Ising SOC in gated TMDs can jointly amplify the Nernst signal, which is highly tunable in nature by external gating. The generated Nernst signal is shown to enhance by at least 1-2 orders of magnitude compared to the Nernst signal generated from a purely intrinsic Ising SOC. 
We first calculate the Berry curvature due to Rashba SOC, and show that the Berry curvature is strongly enhanced and sharply peaked in the momentum space around the valley points, resembling a Dirac-delta distribution. We then calculate the Nernst response and show that even in the absence of Rashba SOC the Nernst signal at a valley point is shown to be purely dominated by the anomalous geometrical contribution, and the conventional $B-$dependent Nernst signal is significantly weaker for typical material parameters. 
The tunable anomalous valley Nernst signal in the presence of Rashba coupling is shown increase by at least one order of magnitude compared to the Nernst signal generated from a purely intrinsic Ising SOC. Although the total Nernst signal, i.e. summing the contributions from both valleys, vanishes due to preservation of time reversal (TR) symmetry, a small external $B-$field lifts the valley degeneracy and further amplifies the Nernst signal by almost two orders of magnitude. 

The Nernst effect in bilayer TMDs also has received limited attention so far. Additionally, here we briefly discuss the band structure, Berry curvature of bilayer TMDs~\cite{Kormanyos:2018}, and theoretically examine the generated Nernst signal due to large Berry curvature in these materials. Particularly, we calculate the Nernst response of a 3R and 2H stacked bilayer MoS$_2$. Including the intrinsic SOC effects, the 3R stacked bilayer MoS$_2$ breaks inversion symmetry, unlike the 2H stacked bilayer which remains spin-degenerate. A finite Rashba SOC can enhance and tune the valley Berry curvature of a 3R stacked bilayer. On the other hand a small Rashba SOC generates (and can also tune) a giant Berry curvature in a 2H stacked bilayer due to lifting of spin-degeneracy. We calculate the anomalous Nernst response and show that Rashba SOC can also enhance and tune the Nernst signal in bilayer TMDs.
Our predictions can be directly tested in experiments, and can be employed for several promising caloritronics applications. 

This paper is organized as follows: in Section II we discuss the Hamiltonian and the Berry curvature of monolayer TMDs in the presence of Rashba spin orbit coupling, and show the tunability of the Berry curvature with the Rasba parameter. In Section III we discuss the Nernst effect in monolayer TMDs, showing the enhancement and modulation of the anomalous Nernst coefficient with Rashba SOC. In Section IV, we discuss the band structure, Berry curvature and Nernst effect in bilayer TMDs focusing on 3R stacked bilayer MoS$_2$ and 2H stacked bilayer MoS$_2$. We then also discuss the enhancement and tunability of the anomalous Nernst coefficient in both these cases. 
Finally we summarize and conclude in Section V. 

\begin{figure}
	\includegraphics[scale=0.21]{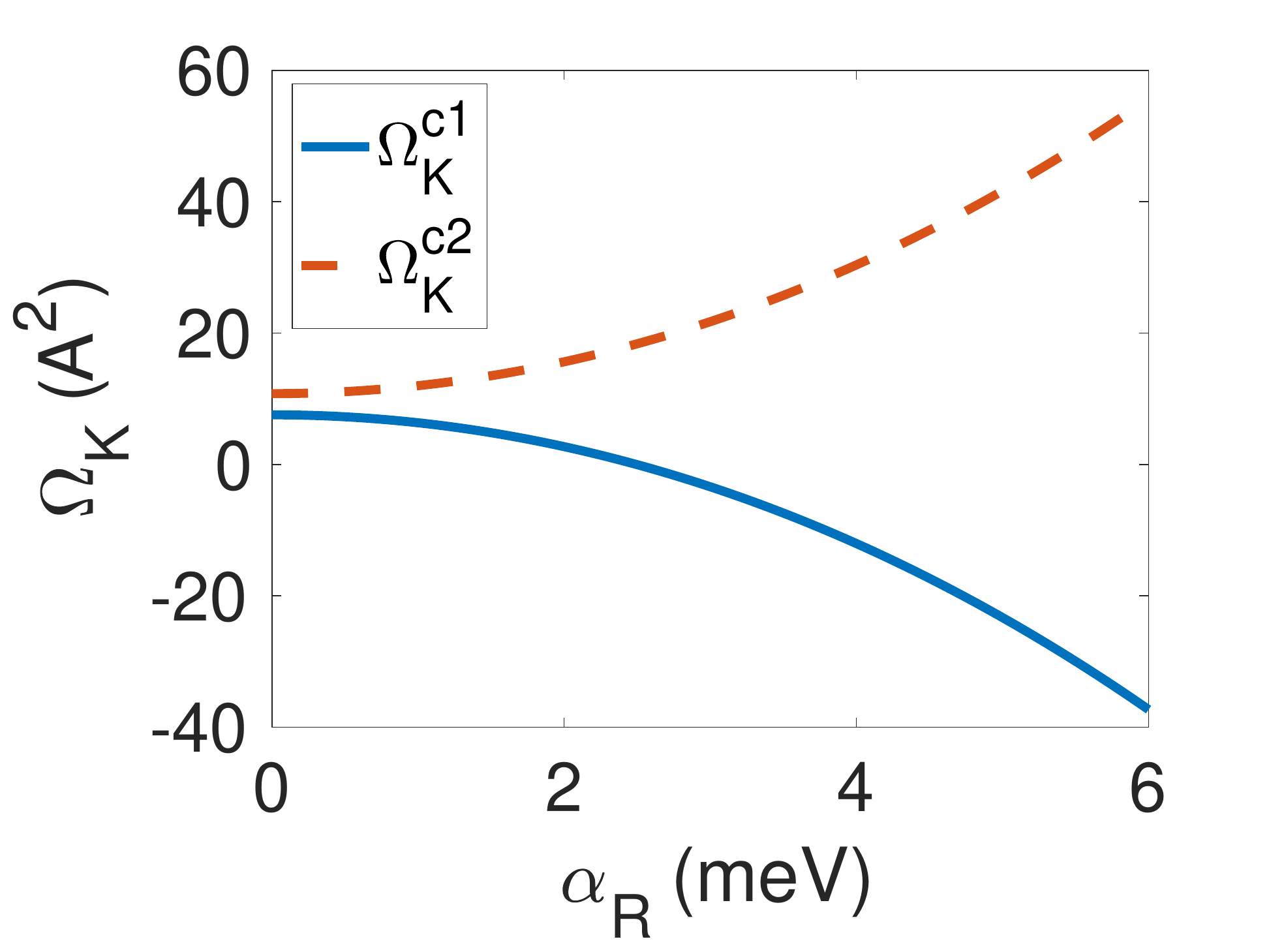}
	\includegraphics[scale=0.21]{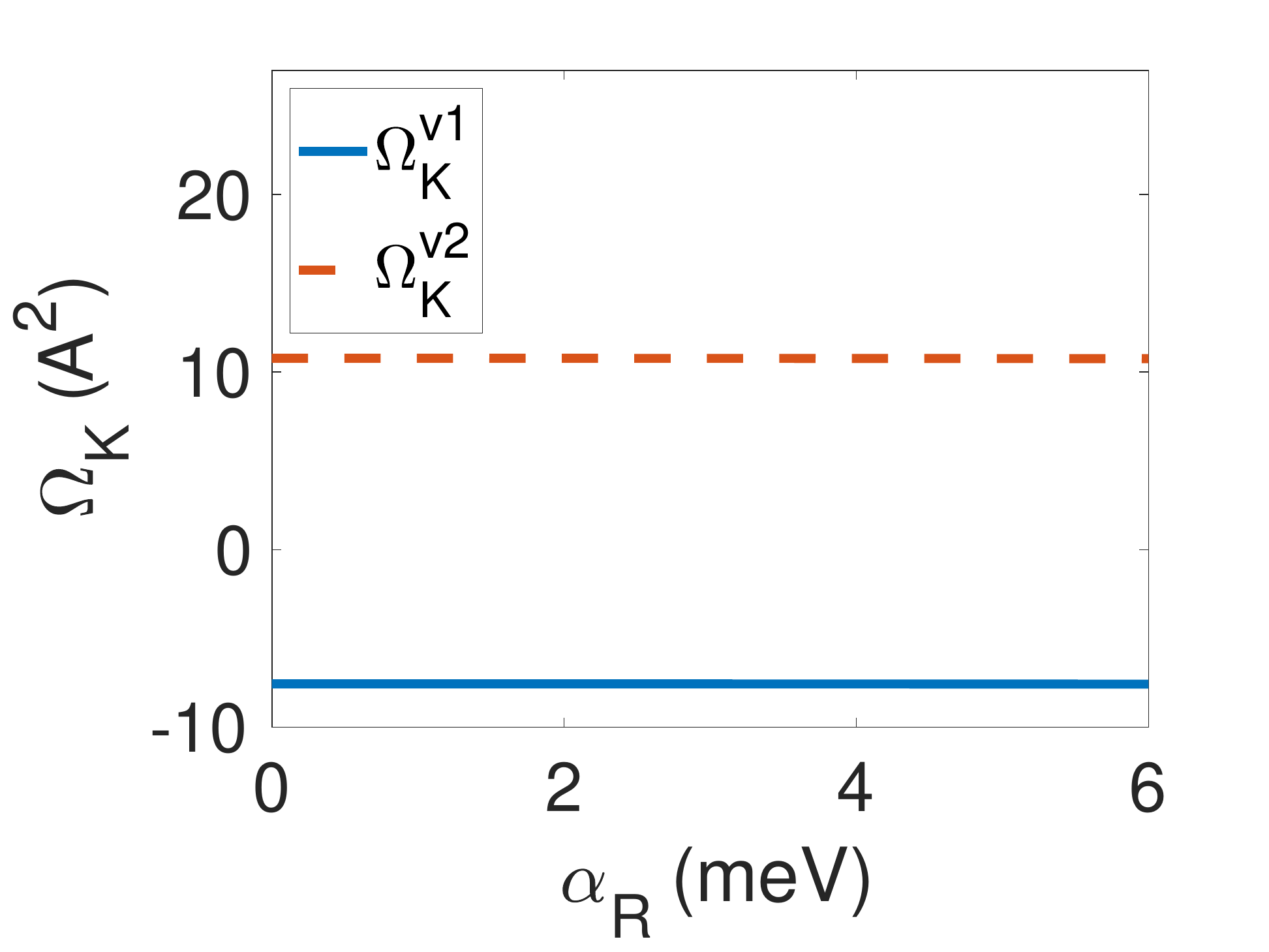}
	\includegraphics[scale=0.11]{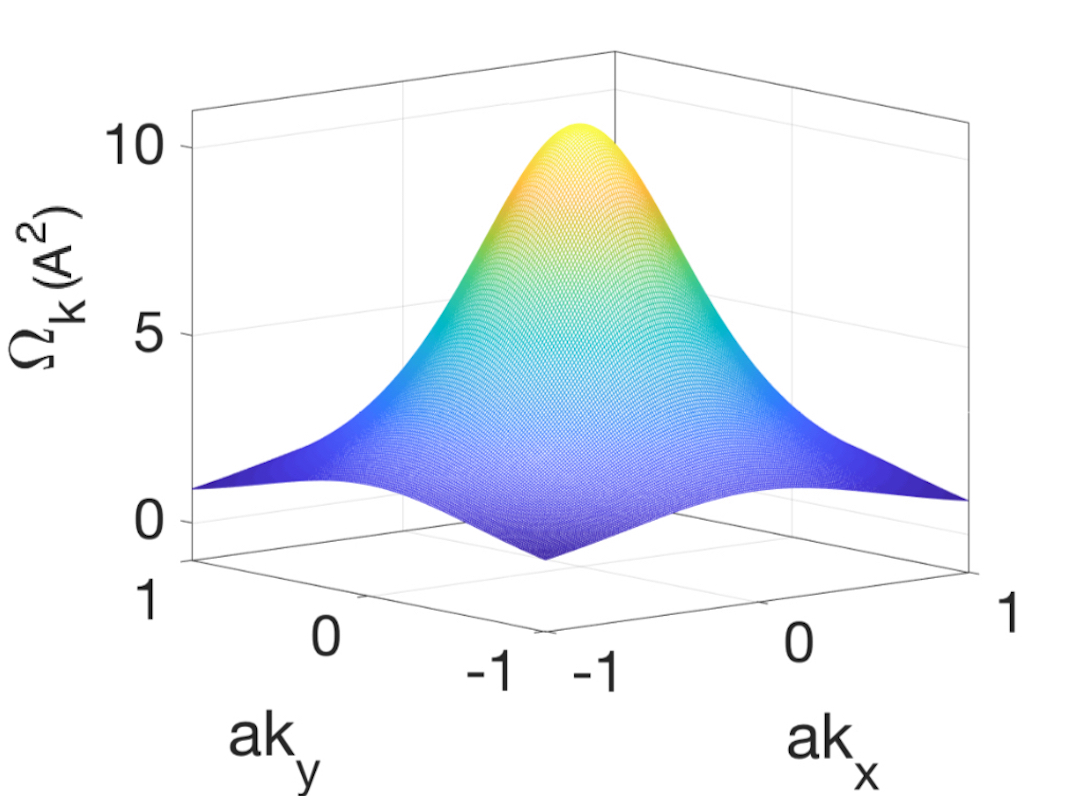}
	\includegraphics[scale=0.11]{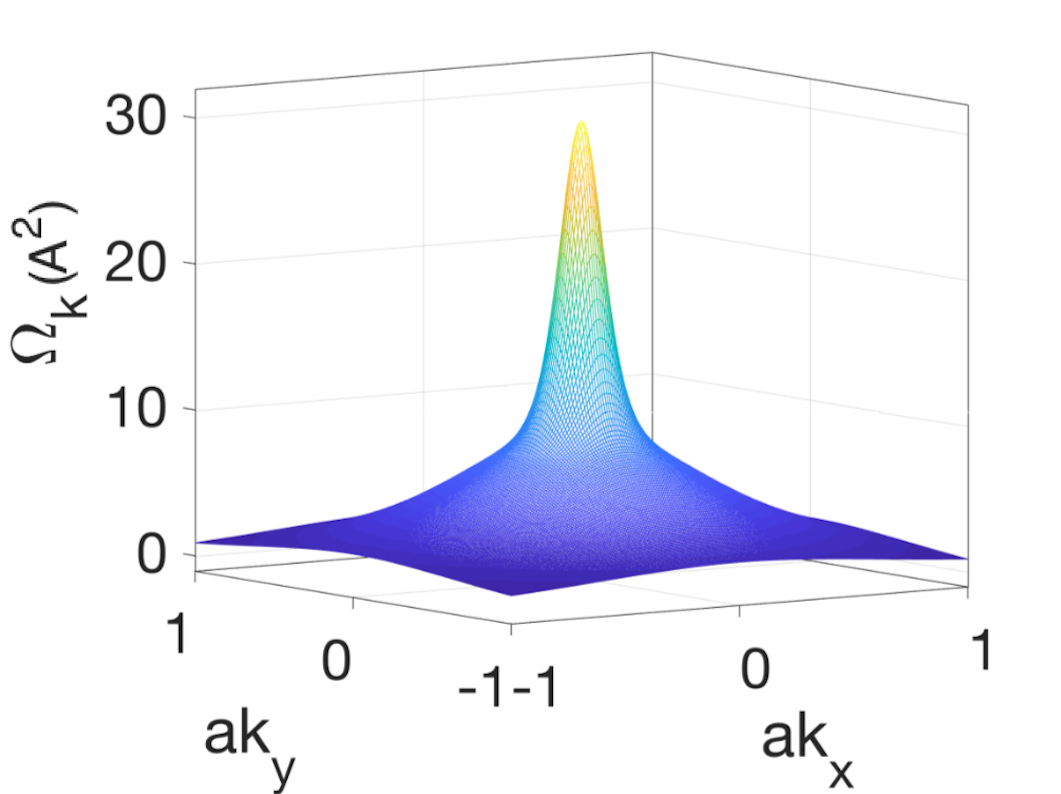}
	\caption{(color online) Tunable Berry curvature in TMDs. \textit{Top Left:} The Berry curvature of the conduction bands can be tuned by varying $\alpha_R$. \textit{Top Right:} The valence bands do not show significant variation with $\alpha_R$ becasue $\alpha_R\ll l_v$.  \textit{Bottom Left:} Berry curvature near the valley point for $\alpha_R=0$. \textit{Bottom Right:} Berry curvature near the valley point for $\alpha_R=4meV$ displaying a higher magnitude and a much a sharper peak at the valley point $\mathbf{K}$. The parameters here are chosen for MoS$_2$.}
	\label{Fig_Berry_curv1}
\end{figure}

\section{Hamiltonian and Berry curvature}
We begin with the low energy $\mathbf{k}\cdot\mathbf{p}$ Hamiltonian for a generic monolayer TMD, which is given by~\cite{Xiao:2012} 
\begin{align}
H_0 = at (\tau k_x \sigma_x+ k_y\sigma_y) + \frac{\Delta}{2}\sigma_z
\end{align}
where $a$ is the lattice constant, $t$ is the effective hopping integral, $\Delta$ is the bare energy bandgap, $\mathbf{k}$ is the momentum measured from the valley point $\mathbf{K}$, and $\sigma_i's$ represent Pauli matrices for the electron bands. The intrinsic spin orbit coupling can be represented by the following Hamiltonian~\cite{Xiao:2012} 
\begin{align}
H_{SO} = \lambda_c \tau \frac{\sigma_0+\sigma_z}{2} s_z + \lambda_v\tau\frac{\sigma_0 - \sigma_z}{2} s_z
\end{align}
where $\lambda_c$ ($\lambda_v$) is the intrinsic Ising spin-orbit coupling in the conduction (valence) band, $s_i^s$ represent Pauli matrices for the electron spin, and $\tau\in\{+1,-1\}$ is the binary valley index. This intrinsic SOC lifts the spin-degeneracy and the bands become spin-polarized.
In addition to the intrinsic spin orbit coupling, a Bychkov-Rashba coupling can also arise in TMDs broken out-of-plane mirror symmetry due to external gating. In the context of the $\mathbf{k}\cdot\mathbf{p}$ Hamiltonian this term can be written as~\cite{Ochoa:2013} 
\begin{align}
H_R = \alpha_R (\tau \sigma_x s_y - \sigma_y s_x)
\end{align}
where $\alpha_R$ is the strength of the Bychkov-Rashba SOC. The total Hamiltonian is thus given by
\begin{align}
H = H_0 + H_{SO} + H_R
\label{Eq_H_tot}
\end{align}
The energy eigenvalues of the Hamiltonian $H$ specifically at the valley points (i.e. $\mathbf{k}=0$) are 
\begin{align}
&E_{n\tau} = n \frac{\lambda_c \tau}{2} - n\frac{\lambda_v \tau}{2} \nonumber \\ 
&\pm \frac{1}{2} \sqrt{(\lambda_c + \lambda_v)^2 + \Delta^2 + 2n\Delta\tau (\lambda_c+\lambda_v) + 8\alpha_R^2 (1-n\tau)}
\end{align}
where $n\in\{+1,-1\}$. 
The spin degeneracy in the conduction bands is broken as a result of $\lambda_c$ and $\alpha_R$. Unlike $\lambda_c$ which is an intrinsic SOC, the Rashba SOC allows us to tune the energy bands by external electrostatic gating. 

The presence of Berry curvature in TMDs can result in anomalous transport responses such as the anomalous Hall, spin Hall and the Nernst effects~\cite{Niu:2006}. 
The Berry curvature for a generic $m^{th}$ band is given by the expression
\begin{align}
\Omega_{ab}^m = i\sum\limits_{m\neq m'}\frac{\langle m|\partial H/\partial k_a|m'\rangle \langle m'|\partial H/\partial k_b|m\rangle -(a\leftrightarrow b )}{(E_m' - E_m)^2} \nonumber \\
\label{Eq_Berry_curv}
\end{align}
The presence of a finite Rashba SOC can thus allow us to manipulate the Berry curvature of the electron bands. Even though the Berry curvature for the Hamiltonian $H$ in Eq.~\ref{Eq_H_tot} can be evaluated analytically, the exact expressions are cumbersome and we do not provide them here. Instead, we directly numerically calculate the Berry curvature and related quantities. 
Fig.~\ref{Fig_Berry_curv1} shows the calculated Berry curvatures of the conduction bands as a result of tuning the Rashba SOC parameter $\alpha_R$. We note that the Berry curvature of the conduction bands can increase up to atleast 1 order of magnitude greater than the valence bands for typical values of $\alpha_R$. On the other hand, the valence band Berry curvature stays close to the orbital (intrinsic) Berry curvature ($\Omega_{\mathbf{K}} = -2a^2t^2 / (\Delta-\lambda_v)^2$) as the intrinsic SOC $\lambda_v \gg \alpha_R$ typically for all TMDs. The spin splitting $\lambda_c\sim\alpha_R$ and the Berry curvature in conduction bands is strongly affected by both parameters, $\lambda_v$ and $\alpha_R$. Interestingly, we also note that Rashba SOC not only enhances the Berry curvature contribution for the conduction band but also sharpens the peak at the valley point as demonstrated in Fig.~\ref{Fig_Berry_curv1}. The intrinsic Berry curvature scales as $\sim 1/(1+k^2)$ around $\mathbf{K}$ but the Rashba mediated Berry curvature has a much sharper distribution, closely resembling a Dirac-delta function. This is expected because the Chern number of the band is a topological invariant, and therefore a sharp peak in the Berry curvature must be compensated by reducing contributions from  the neighboring points in momentum space.

\section{Nernst effect in monolayer TMDs}
 Nernst effect refers to the generation of a transverse electric field in the presence of a longitudinal temperature gradient. Conventionally, the Nernst effect should occur only in the presence of an external magnetic field, which provides a transverse velocity to the electrons by the Lorentz force. However, a non-trivial Berry curvature $\mathbf{\Omega}_\mathbf{k}$ of the bands can also give rise to a Nernst response as a result of an anomalous velocity term~\cite{Niu:2006} genrated by $\mathbf{\Omega}_\mathbf{k}$. 
A non-zero Nernst response is expected only when overall time-reversal symmetry is violated. In doped TMDs the contribution from the spin-up and spin-down bands exactly cancels out each other as the system preserves time-reversal symmetry. 
The anomalous Nernst coefficient is given by the product of the anomalous velocity and the entropy density~\cite{Tewari:2008} 
\begin{align}
\alpha_{xy} = \frac{k_Be}{\hbar} \sum\limits_{m=1}^{4}{\int d^2\mathbf{k } \Omega^m_{\mathbf{k}} s^m_{\mathbf{k}}}
\label{Eq_alphaxy_anom}
\end{align}
where $m$ is the band index, $s_{\mathbf{k}} = -n_f \log n_f - (1-n_f) \log (1-n_f)$ is the entropy density, $n_f$ being the Fermi distribution function. It is important to point out that the anomalous Hall conductivity $\sigma_{xy}$ directly depends on the Berry curvature of the filled bands, but on the other hand $\alpha_{xy}$ is a Fermi surface quantity, because $s_{\mathbf{k}}$ is zero for completely filled and completely empty bands. 
\begin{figure}
\includegraphics[scale=0.21]{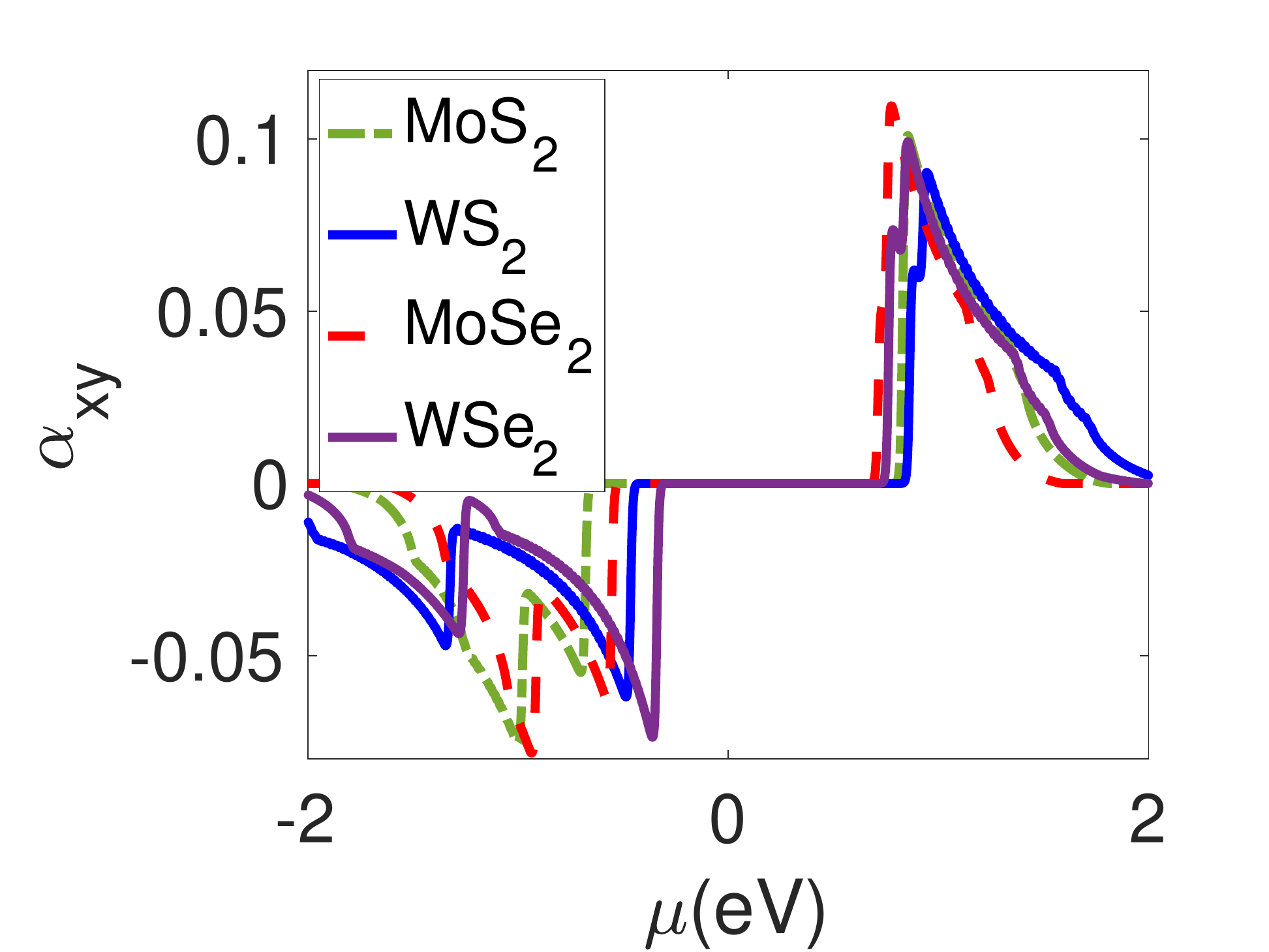}
\includegraphics[scale=0.21]{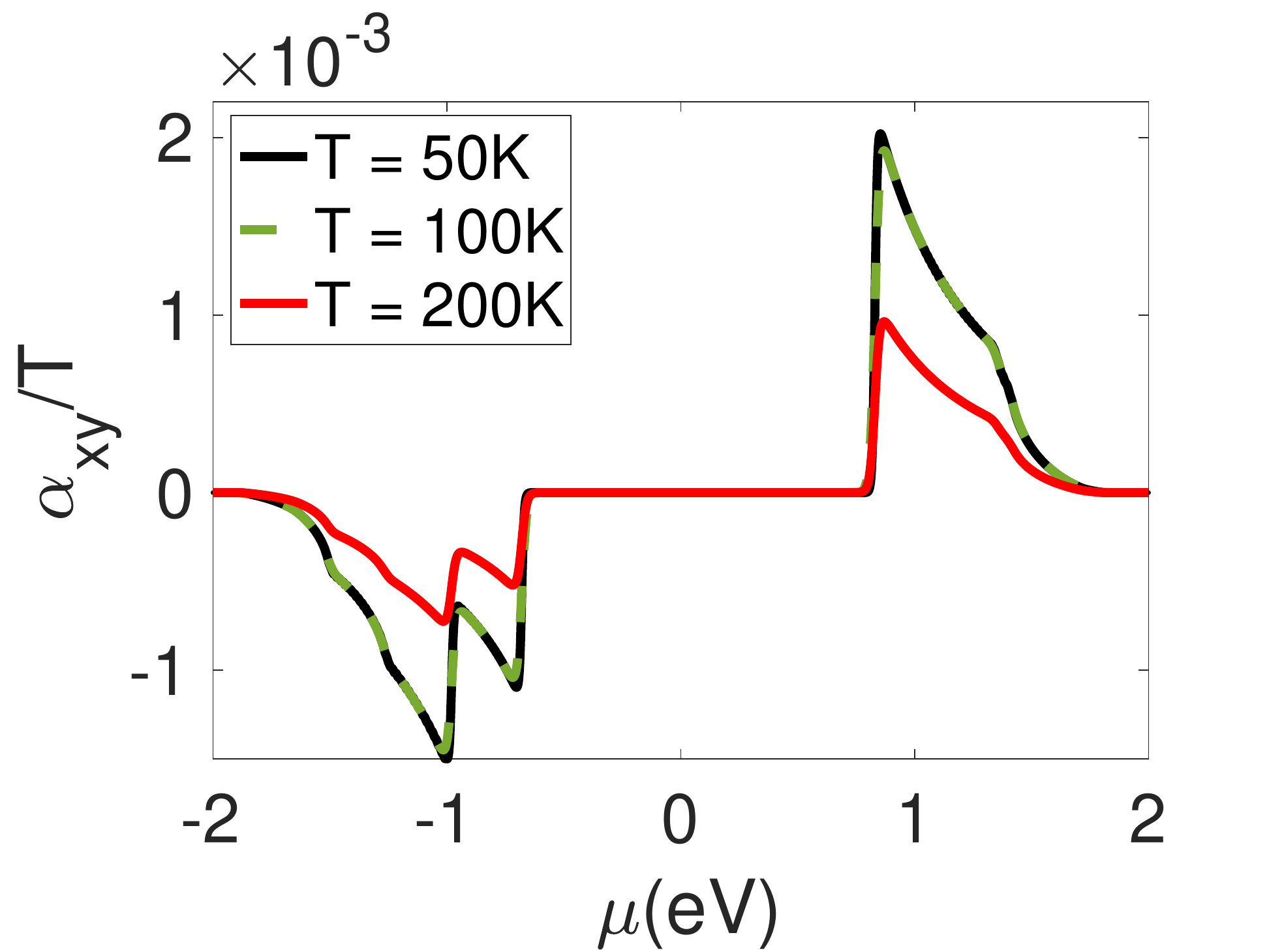}
\includegraphics[scale=0.21]{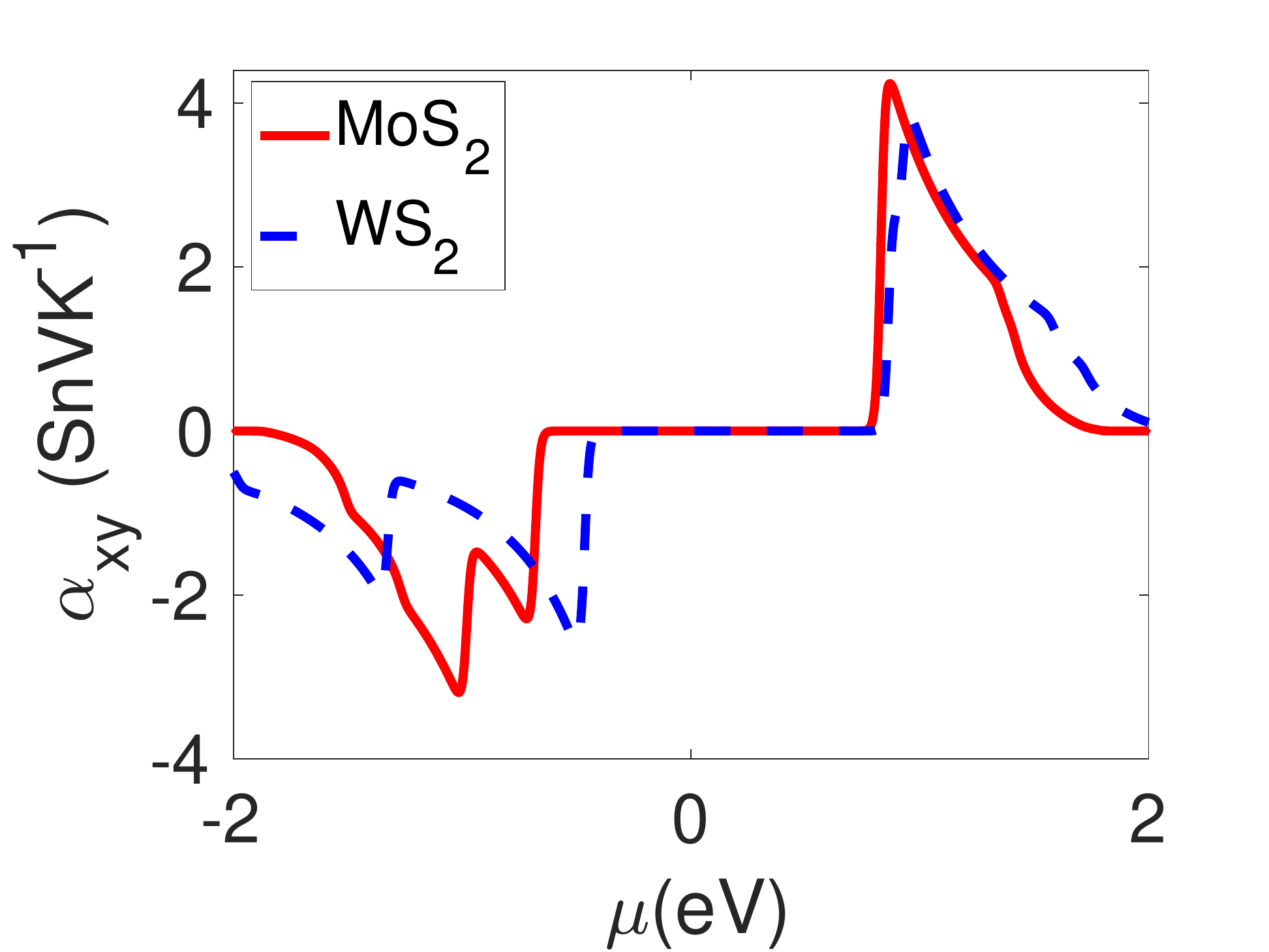}
\includegraphics[scale=0.21]{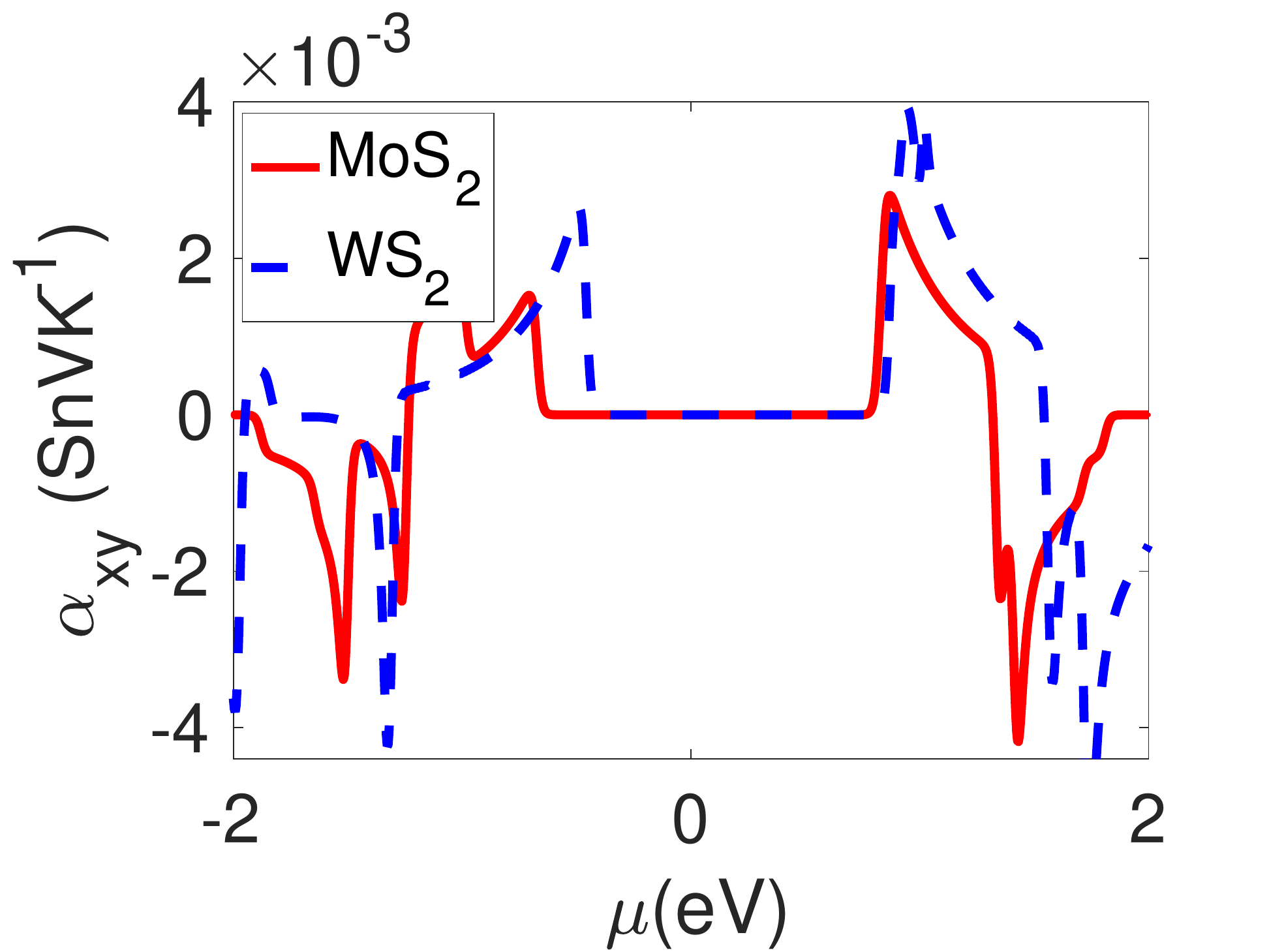}
\caption{(color online) \textit{Top Left:} Anomalous Nernst coefficient (in the units $k_B e/\hbar$) at the valley $\tau=1$ as a function of chemical potential $\mu$ for various TMDs at $T=50K$. \textit{Top Right:} Anomalous Nernst coefficient (in the units $k_B e/\hbar K$) at valley $\tau=1$ for MoS$_2$ at various temperatures. The Rashba SOC parameter was chosen to be $\alpha_R = 0$. \textit{Bottom left:} Anomalous Nernst coefficient (units $SnVK^{-1}$) at $\tau=1$, $\alpha_R=0$ and $T=100K$. \textit{Bottom right:} Conventional $B-$dependent Nernst coefficient (units $SnVK^{-1}$) at $\tau=1$ at $T=100K$ and $B = 3T$. The contrast in the magnitudes can be noted.}
\label{Fig_alphaxy_1}
\end{figure}
\begin{figure}
	\includegraphics[scale=0.33]{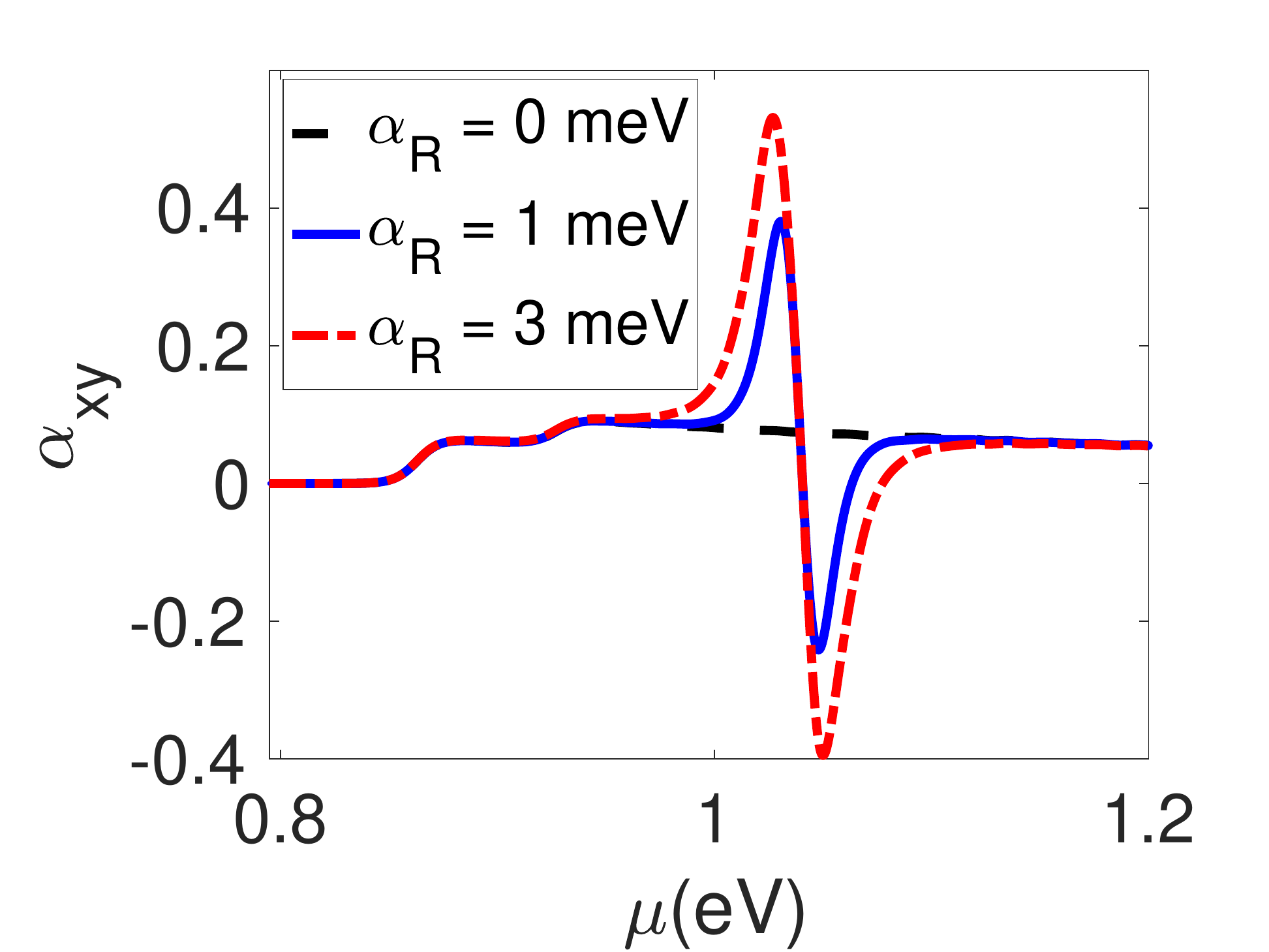}
	\includegraphics[scale=0.18]{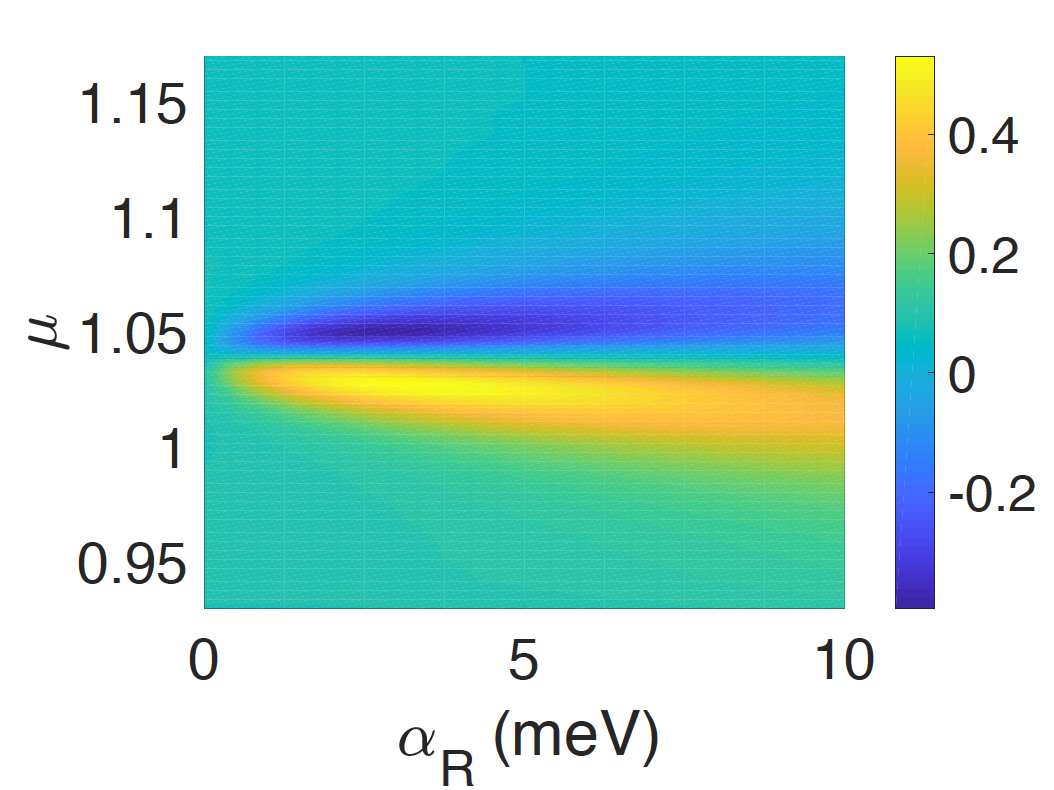}
	\caption{(color online) \textit{Top:} Anomalous Nernst coefficient (in the units $k_B e/\hbar$) at valley $\tau=+1$ as a function of the chemical potential $\mu$ calculated for WS$_2$. For $\alpha_R>0$ the Nernst coefficient shows a significantly enhanced signal compared (by at least one order of magnitude) to the Nernst coefficient contribution from only intrinsic SOC. \textit{Bottom:} Density plot for the anomalous Nernst coefficient for various values of $\alpha_R$ and $\mu$. The sharp peaks when $\mu$ is tuned to partially fill up one of the conduction bands is a striking feature which emerges from a finite Rashba SOC. } 
	\label{Fig_alphaxy_2}
\end{figure}
It is for this reason, that an insulator can give rise to an anomalous Hall response ($\sigma_{xy}$), but not an anomalous Nernst response $\alpha_{xy}$. %Unlike the Hall signal, the Nernst signal also does not directly correlate with the domiant charge carrier type in the system. 
Typically in Fermi liquids the conventional Nernst effect is expected to be small due to Sondheimer's cancellation~\cite{Sondheimer}, which breakdowns of for Berry mediated Nernst effects, as shown recently~\cite{Sharma3}.

\begin{figure}
	\includegraphics[scale=0.1]{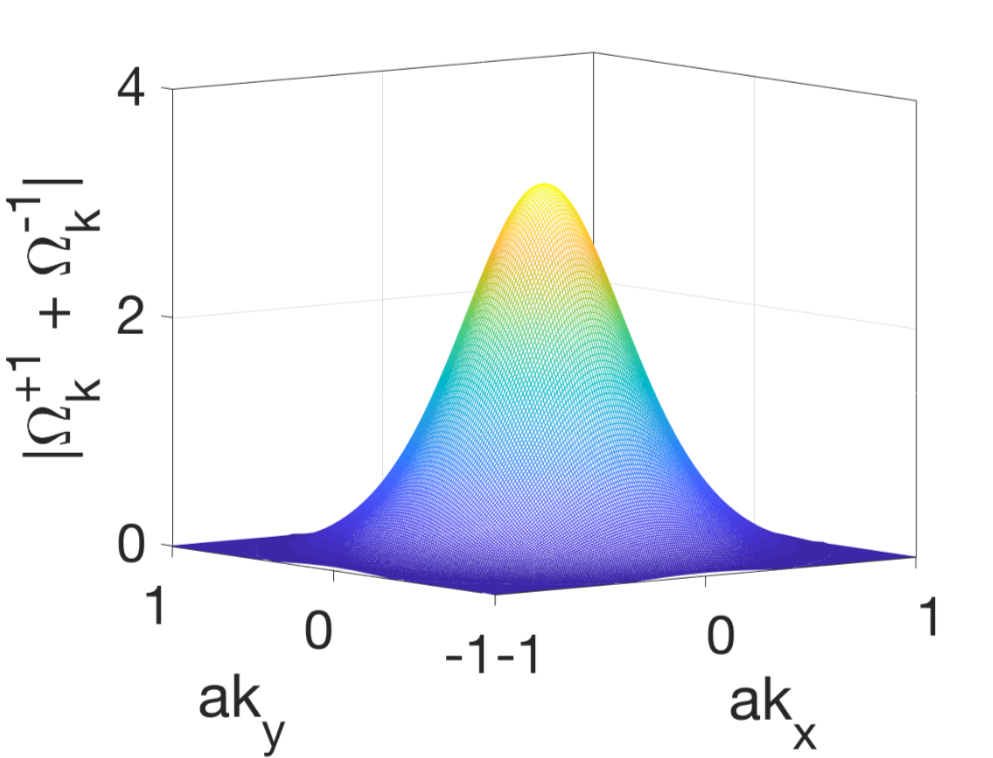}
	\includegraphics[scale=0.1]{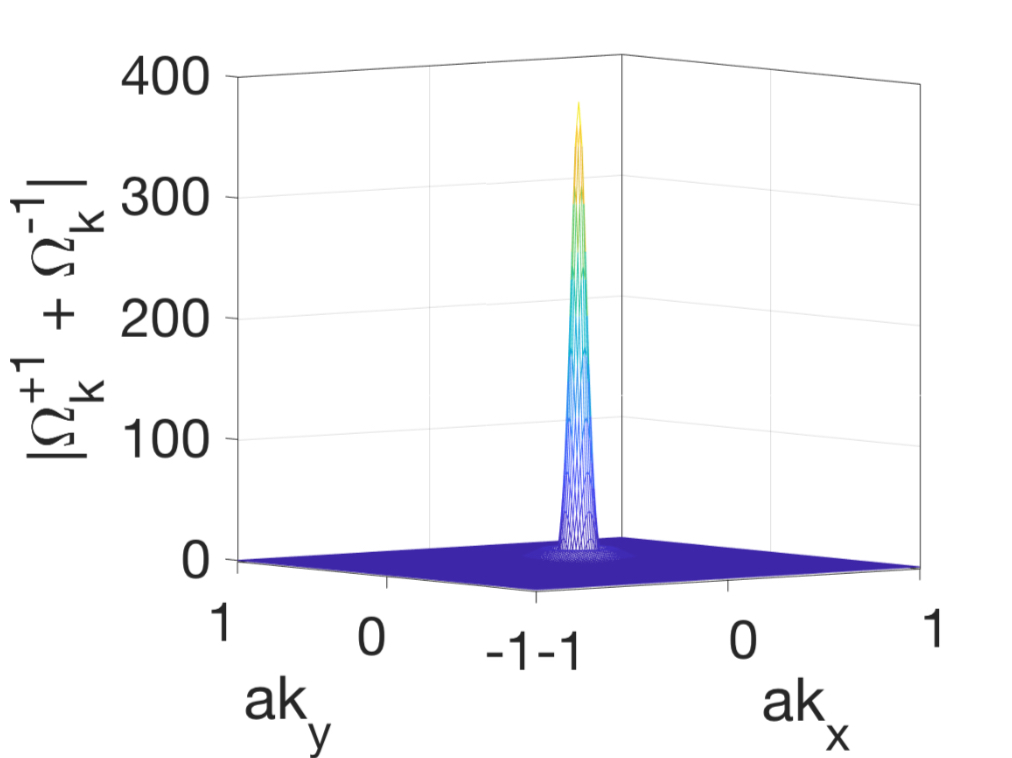}
	\includegraphics[scale=0.17]{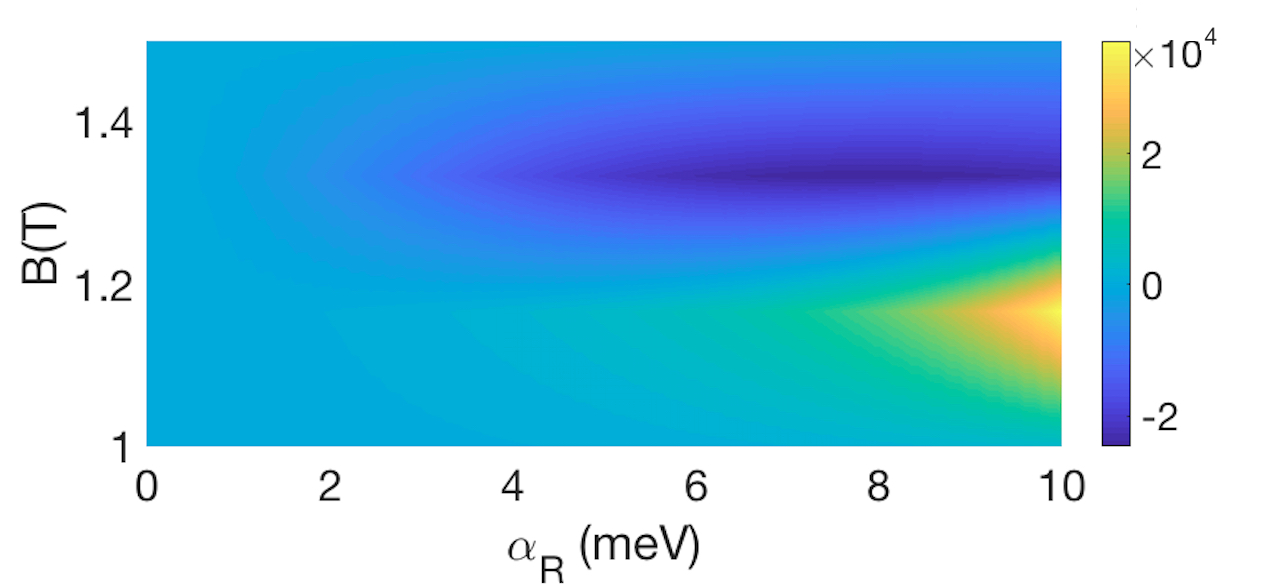}
	\caption{(color online) \textit{Top Left:} The magnitude of the sum of Berry curvatures (in units $\AA^2$) in the two valleys for a particular conduction band in MoS$_2$ when $B=15T$ and $\alpha_R=0$. For $B=0$ we do not expect a net Berry curvature due to TR symmetry. \textit{Top Right:} Same as the left plot but for $\alpha_R=4meV$ and $B=1T$. The Berry curvature now shows a much sharper peak around $\mathbf{K}$ which is more than one order of magnitude greater the former case even for a very small $B$ field.  \textit{Bottom:} Density plot of the sum of peak Berry curvature for various $B$ and $\alpha_R$ values.}
	\label{Fig_Berry_2}
\end{figure}
\begin{figure}
	\includegraphics[scale=0.36]{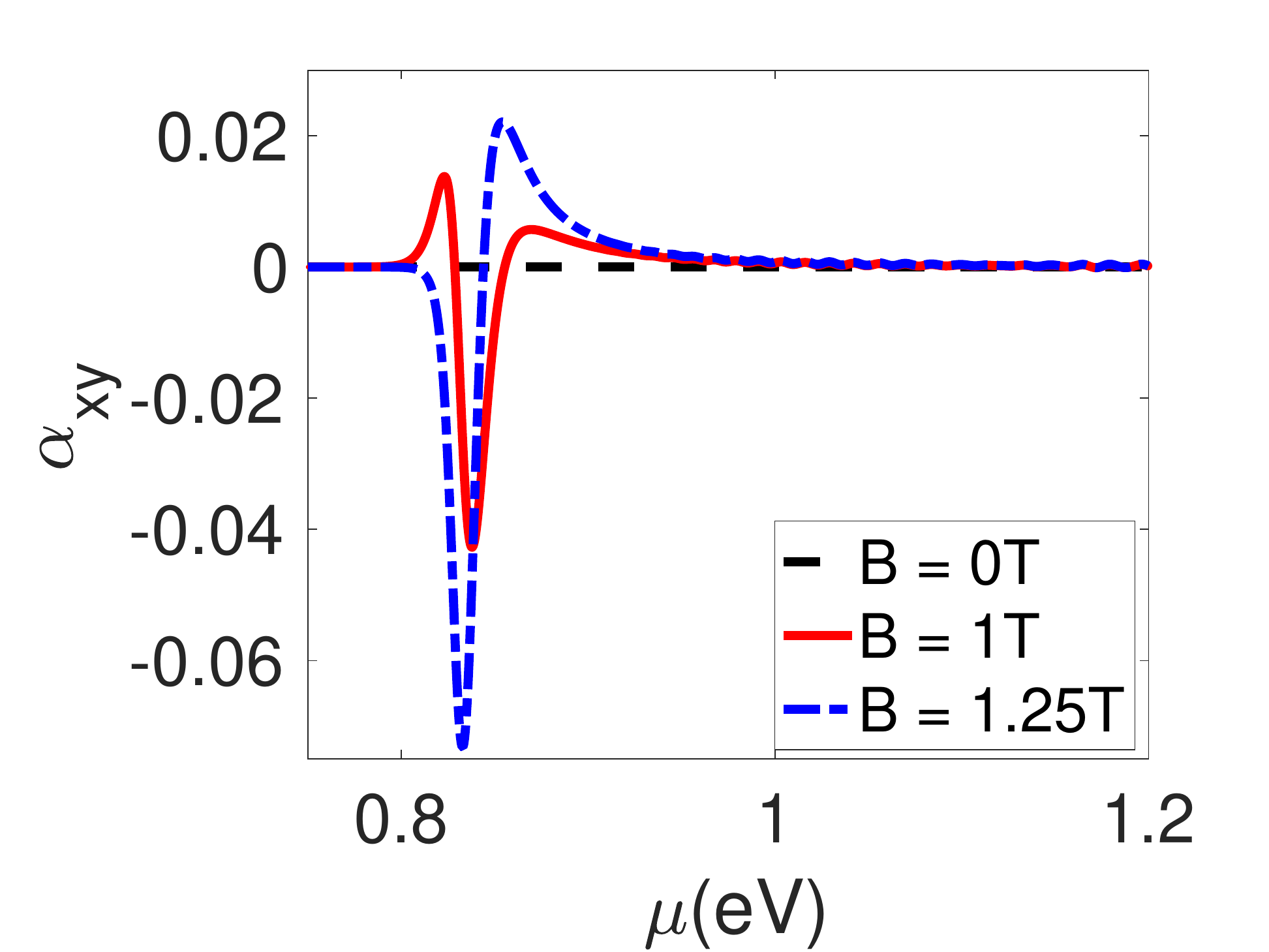}
	\includegraphics[scale=0.36]{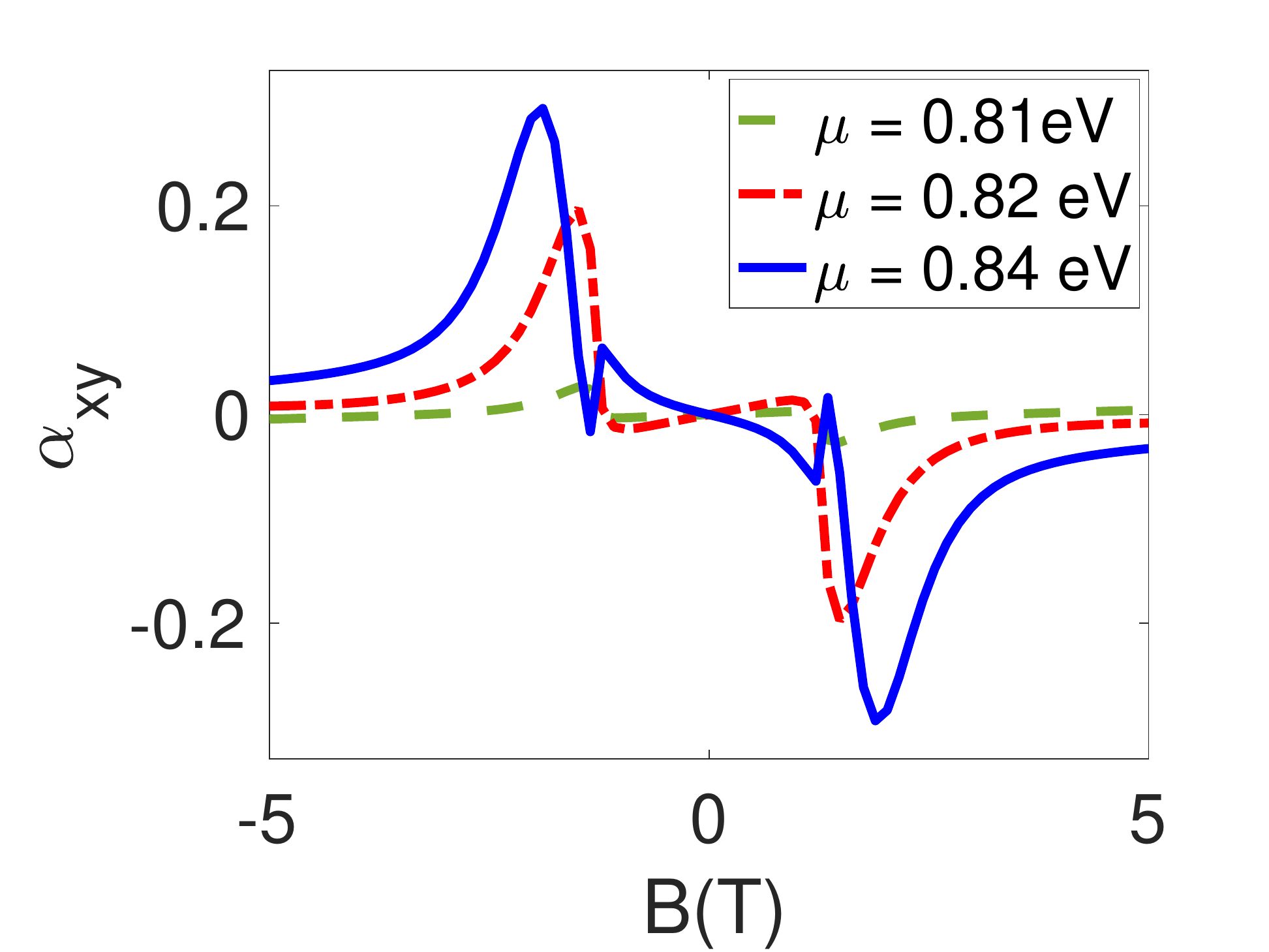}
	\caption{(color online) Total (summing contributions from both valleys) anomalous Nernst coefficient calculated for MoS$_2$. A small $B$-field breaks TR symmetry and the valley symmetry resulting in a large net Nernst response when the chemical potential is tuned appropriately. We chose $T=50K$ and $\alpha_R=4meV$. A large net Nernst signal as a result of a small $B$-field is a distincitive feature in these materials even though the $B-$field by itself is not expected to produce such a large Nernst signal.}
	\label{Fig_totalNernst}
\end{figure}
The anomalous Nernst coefficient is calculated from Eq.~\ref{Eq_alphaxy_anom} by integrating the Berry curvature up to the chemical potential. In Fig.~\ref{Fig_alphaxy_1} we plot the anomalous Nernst coefficient for various TMDs for $\alpha_R =0$. The anomalous Nernst coefficient picks up a non-zero signal for partially filled conduction bands and its magnitude is proportional to the Berry curvature of the respective bands. In the presence of a finite magnetic field TMDs also will exhibit a conventional Nernst response which is given by~\cite{Ziman} 
\begin{align}
\alpha_{xy} = \frac{e^3 \tau_s^2 B}{T\hbar} \int{d^2\mathbf{k} (E - \mu)\left(-\frac{\partial n_f}{\partial E}\right) \left(\frac{ v_x^2\partial^2 E}{\partial^2 k_y^2} - \frac{v_xv_y \partial^2E}{\partial k_x\partial k_y}\right)}
\end{align}  
where $\tau_s$ is the scattering time and $v_i$ is the semiclassical velocity. In order to compare the magnitude of the anomalous part of the Nernst coefficient in TMDs with the conventional one we also calculate the latter within the relaxation time approximation, which is valid in the limit $\omega_c\tau_s\ll 1$, $\omega_c$ being the cyclotron frequency. For typical values~\cite{Brumme} of $\tau_s$ and relevant $B-$ fields we find that this condition is satisfied. Fig.~\ref{Fig_alphaxy_1} shows a comparison of the anomalous and conventional Nernst coefficients. Even when $\alpha_R=0$ the Berry curvature mediated Nernst response is at least 3 orders of magnitude bigger than the conventional Nernst coefficient. 

In Fig.~\ref{Fig_alphaxy_2} we plot the anomalous Nernst coefficient at valley $\tau=+1$ as a function of the chemical potential $\mu$ calculated for WS$_2$. For $\alpha_R>0$ the Nernst coefficient shows a significantly enhanced signal compared (by at least one order of magnitude) to the Nernst coefficient contribution arising just from intrinsic SOC ($\alpha_R=0$). The figure also shows the density plot for the anomalous Nernst coefficient for various values of $\alpha_R$ and $\mu$. The sharp peaks in the Nernst signal when $\mu$ is tuned to partially fill up one of the conduction bands is a striking feature which emerges in these systems from a finite Rashba SOC. As the chemical potential is varied from the bottom of the lower conduction band to the upper conduction band, the Nernst signal changes sign because the sign of the Berry curvature is opposite in the two neighboring bands.

Our discussion so far was focused on the Nernst coefficient contribution from a single valley in TMDs. We noted that the anomalous Nernst contribution is significantly larger than the conventional Nernst coefficient. Further the anomalous contribution can be enhanced by at least one order in magnitude as a result of tuning the Rashba SOC $\alpha_R$, but the conventional Nernst isn't much altered with a finite $\alpha_R$. However, as a result of preservation of TR symmetry the Berry curvature satisfies $\Omega^m_{\mathbf{K}} = -\Omega^m_{-\mathbf{K}}$, and thus the total (summing contributions from both valleys) Berry curvature, and hence the total Nernst coefficient exactly vanishes. Nevertheless, breaking the valley degeneracy can result in a finite flux of the Berry curvature and therefore a finite total Nernst effect. One way to generate such asymmetry is by applying a small external magnetic field perpendicular to the 2D plane. The magnetic field acts on the valley and spin sectors as $H^s_z = b_z s_z$, and $H^v_z = \sum\limits_a m^a b_z$, where $b_z$ is the magnetic field, $s_z$ is the spin-Pauli matrix, and valley orbital magnetic moments $m^a$. The orbital magnetic moment is given by 
\begin{align}
m^a = \frac{ie}{2\hbar} \sum\limits_{a\neq a'}\frac{\langle a|\partial H/\partial k_x|a'\rangle \langle a'|\partial H/\partial k_y|a\rangle +c.c}{E_a - E_a'} \nonumber \\
\label{Eq_Berry_orb}
\end{align}

The terms $H_z^s$ and $H^v_z$ can then be added to the total Hamiltonian in Eq.~\ref{Eq_H_tot}. The role of external magnetic field is to break time reversal symmetry and lift the band degeneracy between $+\mathbf{K}$ and $-\mathbf{K}$. As a result the system now possesses an net flux of the Berry curvature. A conventional Nernst coefficient will also be generated in the process of applying an external $B-$field, however it is of a much smaller magnitude compared with the anomalous contributions. 
In Fig.~\ref{Fig_Berry_2} we plot the magnitude of the sum of Berry curvatures in the two valleys for a particular conduction band in MoS$_2$ for a finite magnetic field. When $B=0$ we expect a vanishing net Berry curvature everywhere. 
We note that for $\alpha_R=0$, we need a large $B-$field in order to generate even a small net Berry curvature. However a small finite $B-$field in conjunction with a finite $\alpha_R$ produces a large net Berry curvature (at least 2 orders of magnitude larger) with a very sharp distribution (resembling a Dirac-delta distribution) around the valley point $\mathbf{K}$, clearly distinct from the $\sim 1/(1+k^2)$ distribution for intrinsic Berry curvature. 

In Fig.~\ref{Fig_totalNernst} we plot the total Nernst coefficient for MoS$_2$ in the presence of a finite magnetic field. In plotting this we integrated the Berry curvature of both valleys up to the chemical potential.
The sharply peaked net Berry curvature distribution around the $\mathbf{K}$ point and a finite Rashba SOC produce an amplified anomalous Nernst coefficient, which can be easily tuned by external gating. It is important to emphasize the role of an external magnetic field, which by itself cannot produce the Nernst signal of this magnitude but is crucial in generating Berry curvature asymmetry between the two valleys. Interestingly, similar behavior for the Nernst response has been recently proposed~\cite{Sharma3} and observed~\cite{Liang:2017} in Dirac semimetals. Even though Dirac semimetals are rather quite distinct from TMDs, the close similarity in their Nernst response is an universal manifestation of underlying non-trivial band topology.

\section{Nernst effect in bilayer TMDs}
So far our discussion was focused on Nernst response in monolayer TMDs. The Nernst response in bilayer TMDs has received limited attention so far. The anomalous Nernst effect in bilayer TMDs is directly related to its Berry curvature properties, which in bilayer TMDs has both inter-layer and intra-layer contributions~\cite{Kormanyos:2018}. Further, the stacking of monolayers also effects the band spectrum and the Berry curvature in bilayer TMDs~\cite{Kormanyos:2018}. Like monolayer TMDs, bilayer TMDs also exhibit a finite Berry curvature, which leads to anomalous transport and optical properties, such as the valley and spin Hall effects. As a prototype, we will discuss (i) 3R stacked bilayer MoS$_2$, and (ii) 2H-stacked bilayer MoS$_2$. As we will examine, both of these materials can exhibit a large Nernst effect of purely anomalous origin, tunable with Rashba spin orbit coupling. Since, we already showed in the previous section that the conventional ($B-$ field dependent) contribution to the Nernst signal is small compared to the anomalous contribution, we will neglect this conventional contribution in our analysis and just focus on the anomalous contribution.

\subsection{3R stacked bilayer} 
We first discuss 3R stacked bilayer MoS$_2$, which break inversion symmetry~\cite{Kormanyos:2018}. Due to inversion symmetry breaking Berry curvature is expected to play an important role in transport properties. Near the $\mathbf{K}$ points, the low energy Hamiltonian up to the lowest order in $k$ can be written as 
\begin{align}
H^{3R}=\left( \begin{array}{cccc}
\epsilon_{cb}^b & \gamma_3 k^+ & \gamma_{cc} k^- & 0 \\
\gamma_3 k^- &   \epsilon_{cb}^v   & 0 & \gamma_{vv} k^- \\
\gamma_{cc} k^+ & 0 & \epsilon_{cb}^t   & \gamma_3 k^+\\
0 &\gamma_{vv} k^+  & \gamma_3 k^- &  \epsilon_{vb}^t
\end{array} \right)
\label{Eq_H_3R}
\end{align}
where the momentum $k^{\pm} = (\tau k_x \pm i k_y)$ is measured from the valley points. The subscripts $vb$ and $cb$ refer to valence and conduction bands respectively, while $t$ and $b$ refer to the top and bottom layer respectively. $\gamma_3$ is the intra-layer coupling between the conduction and the valence band (similar to $at$ in the monolayer case), while $\gamma_{cc}$ and $\gamma_{vv}$ are the inter-layer couplings between top and the bottom layers. From Eq.~\ref{Eq_H_3R} it is clear that the Berry curvature will have both intra-layer ($\gamma_3$ dependent) contributions, and inter-layer ($\gamma_{cc}$ and $\gamma_{vv}$ dependent) contributions. For example, at the $\mathbf{K}$ points, the Berry curvature for the conduction band is given by 
\begin{align}
\Omega_{\mathbf{K}} = \frac{1}{2} \left(\frac{2\gamma_3}{\epsilon_{cb}^b - \epsilon_{vb}^b}\right)^2 \pm \frac{1}{2(\epsilon_{cb}^b-\epsilon_{cb}^v)^2} \gamma_{cc}^2
\end{align} 
for the top/bottom later, and similarly one may derive an expression of the Berry curvature for the valence band. For typical material parameters, we find that the contributions of inter-layer and intra-layer on the Berry curvature are of comparable order of magnitude. The above Hamiltonian does not  contain the spin-orbit coupling effects, which may be accounted for by adding the terms $\Delta_{cb} s_z \tau_z$ and $\Delta_{vb} s_z \tau_z$ to the conduction and the valence bands respectively, where the Pauli matrices $s_z$ and $\tau_z$ act on spin and valley space. Note that $\Delta_{vb}\gg \Delta_{cb}$, as is the case in monolayer TMDs. Spin-orbit coupling splits the spin-degenerate bands in both the conduction and the valence bands, though its effect on the conduction bands is rather small. 

\begin{figure}
	\includegraphics[scale=.33]{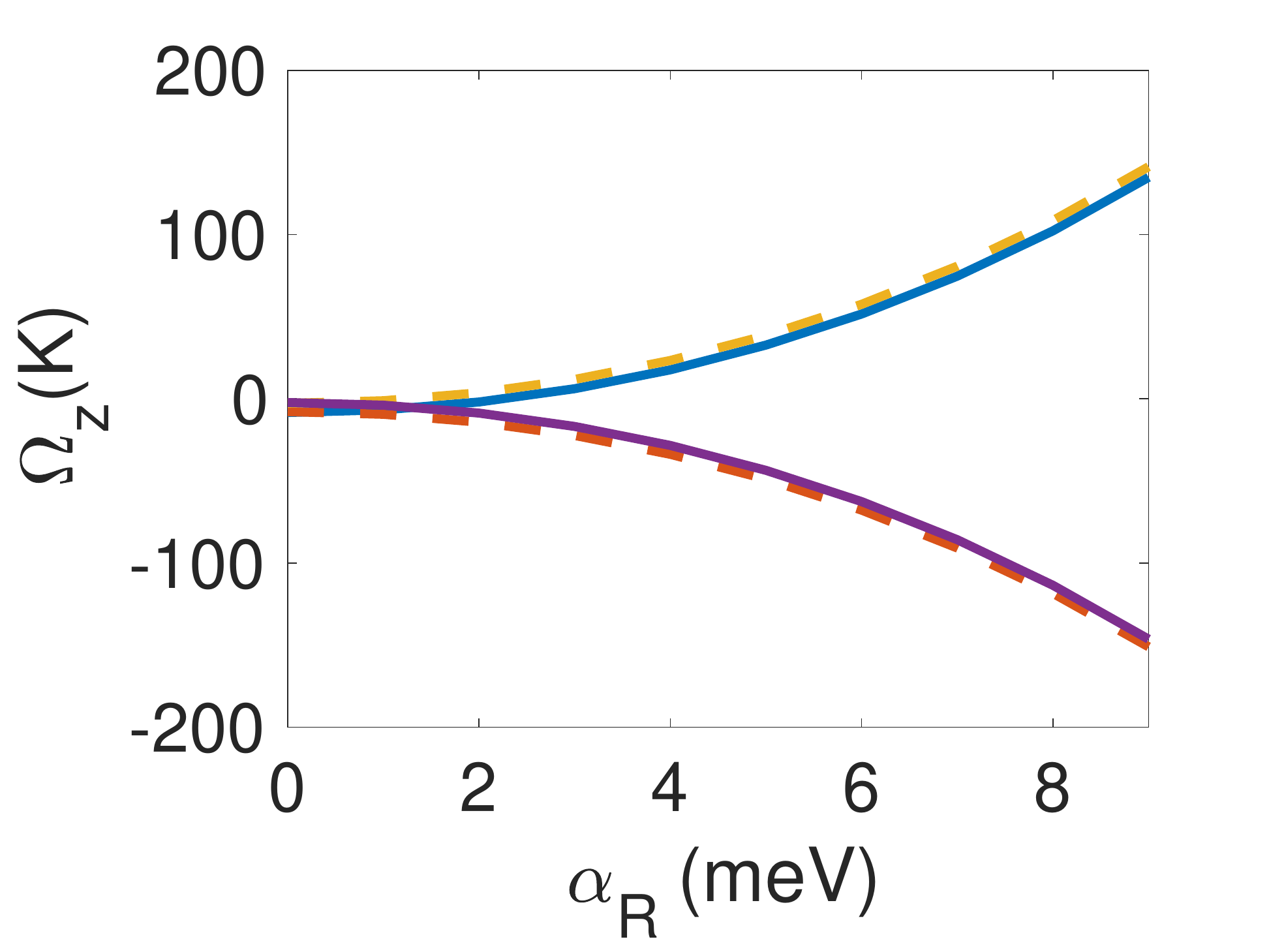}
	\includegraphics[scale=.2]{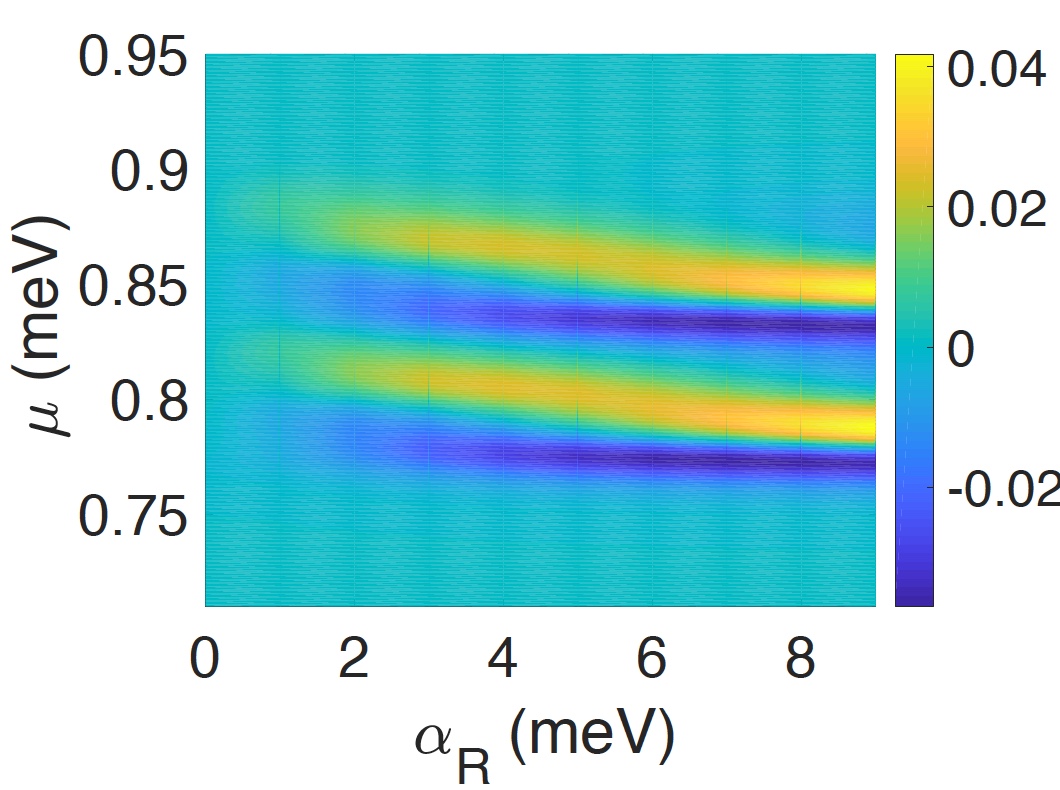}
	\caption{\textit{Top panel:} Berry curvature (in units $\AA^2$) of four out of the eight bands of a 3R stacked bilayer MoS$_2$. The Berry curvature is easily tunable with Rashba spin orbit coupling and can be enhanced up to two orders of magnitude for typical $\alpha_R$ values. \textit{Bottom panel:} The total anomalous Nernst coefficient for a magnetic field of $5T$, tunable with the chemical potential and $\alpha_R$. }
	\label{Fig_bilayer_1}
\end{figure}
\begin{figure}
	\includegraphics[scale=.215]{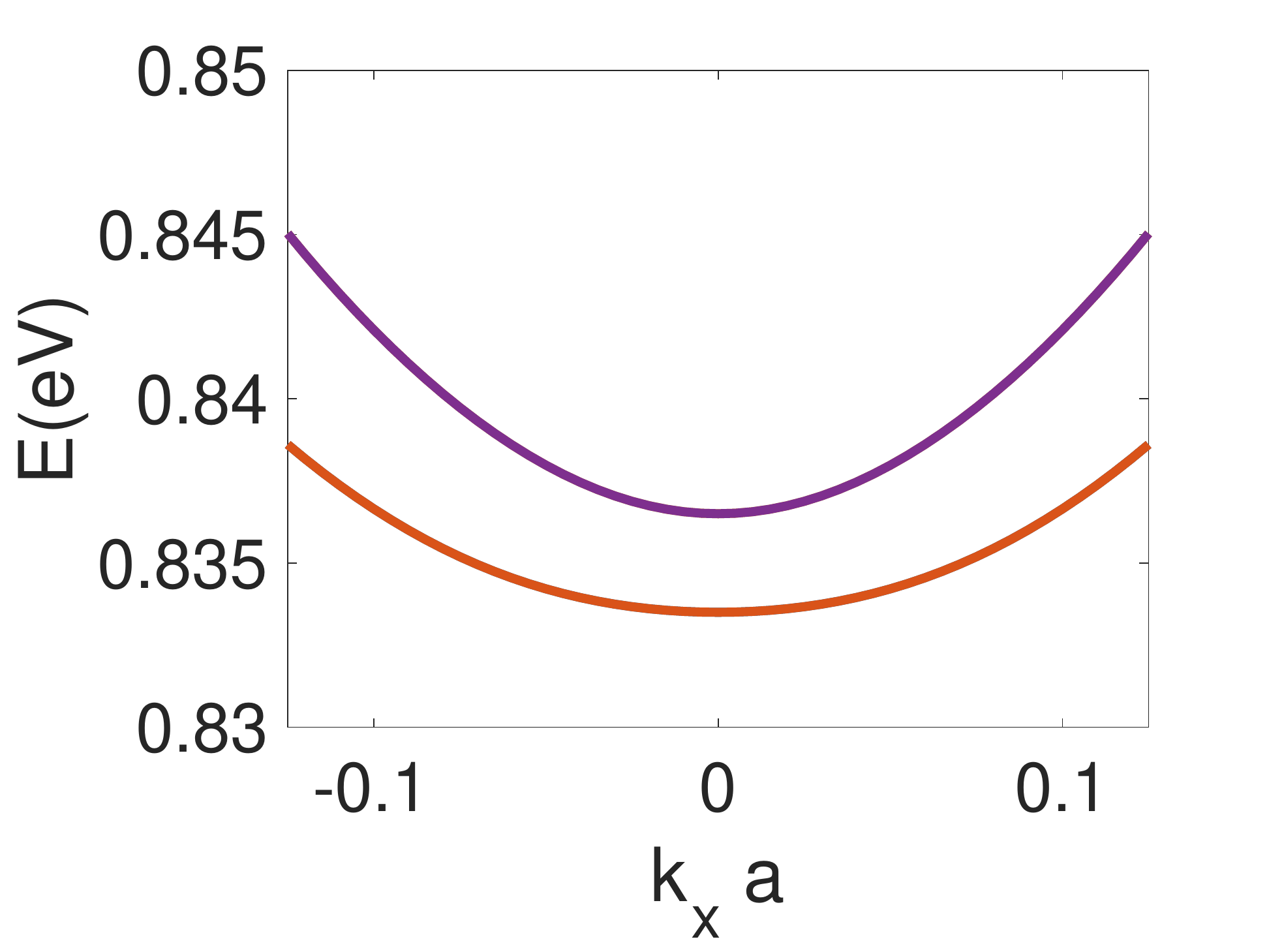}
	\includegraphics[scale=.215]{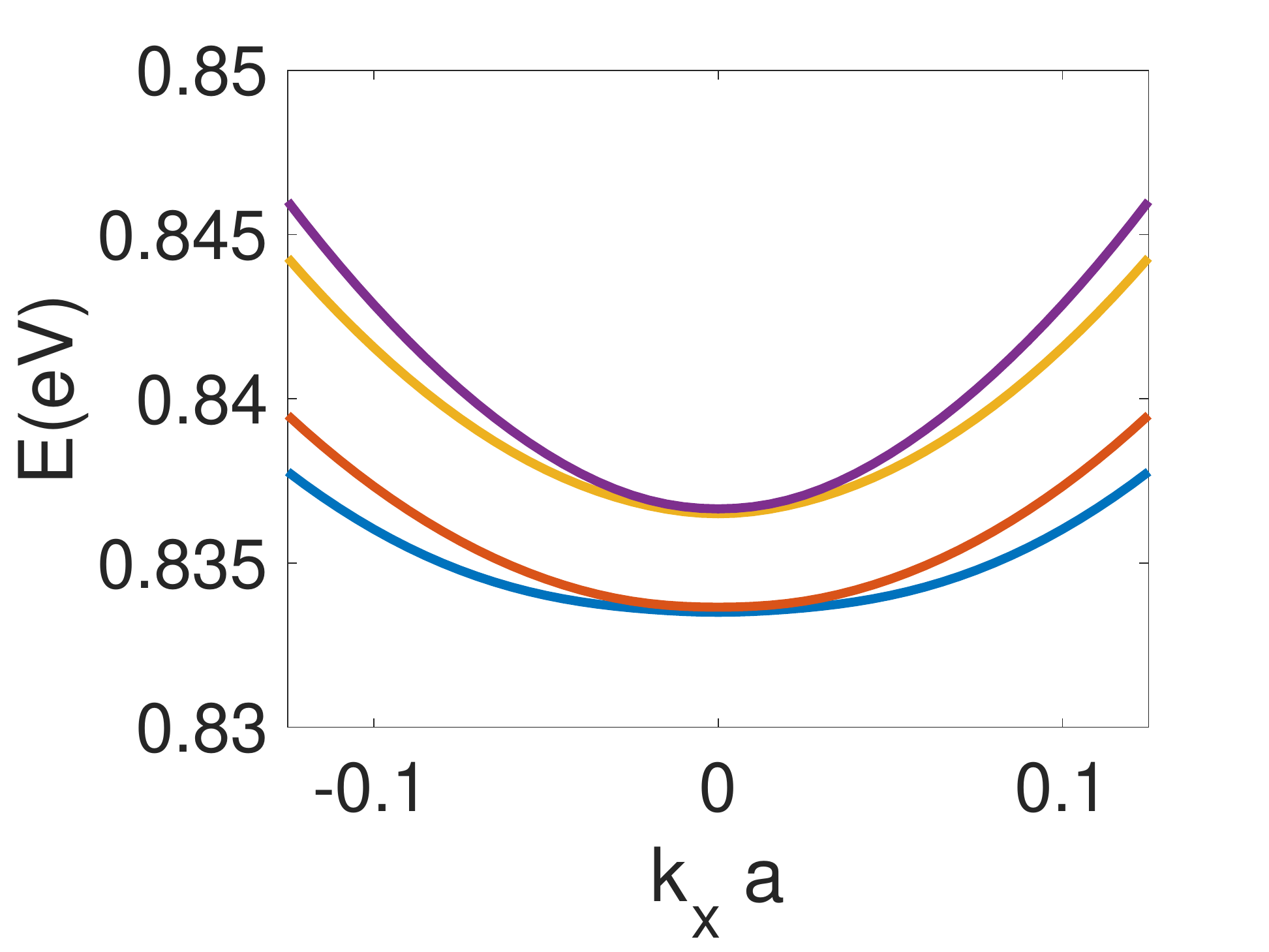}
	\includegraphics[scale=.215]{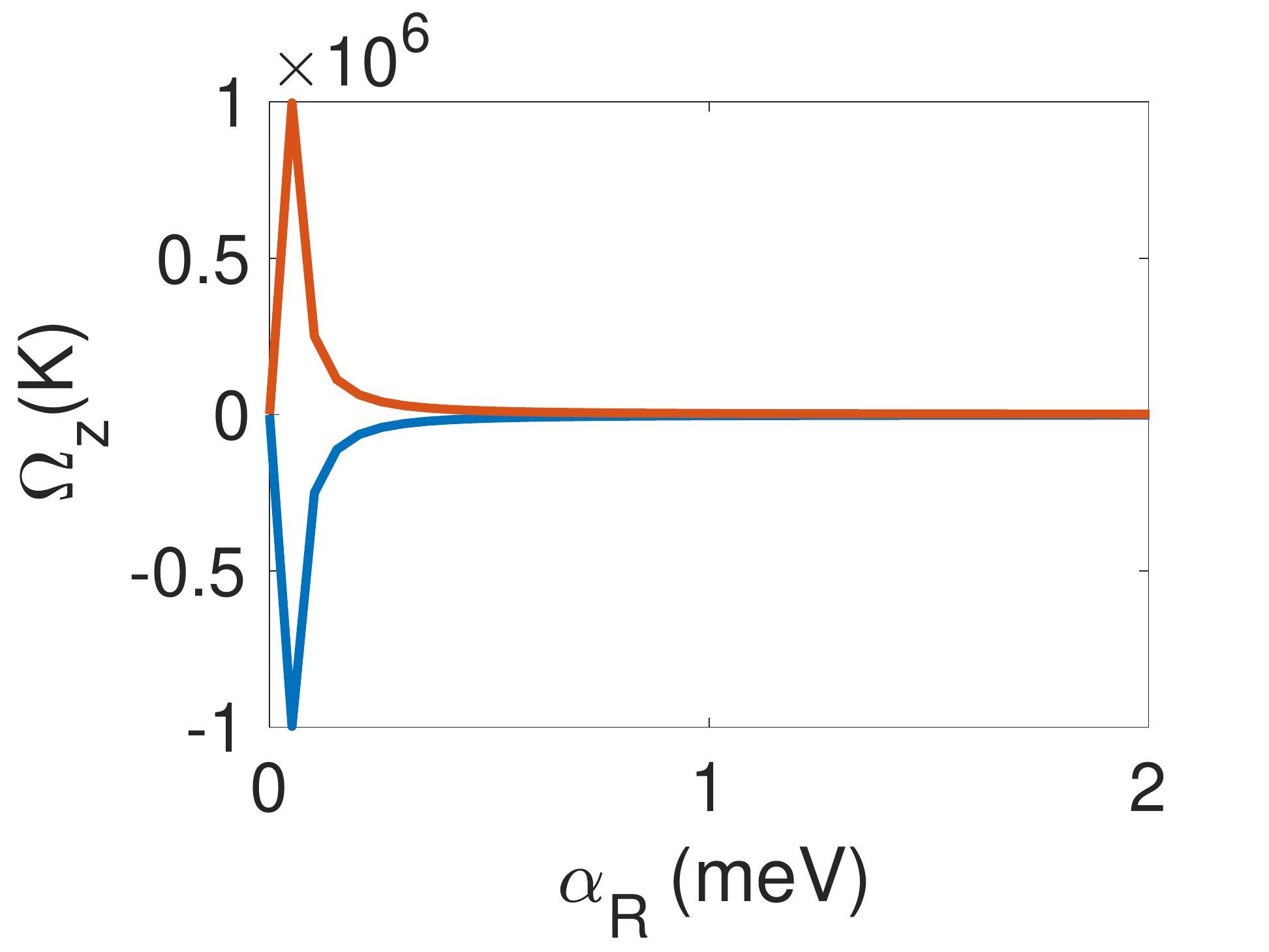}
	\includegraphics[scale=.115]{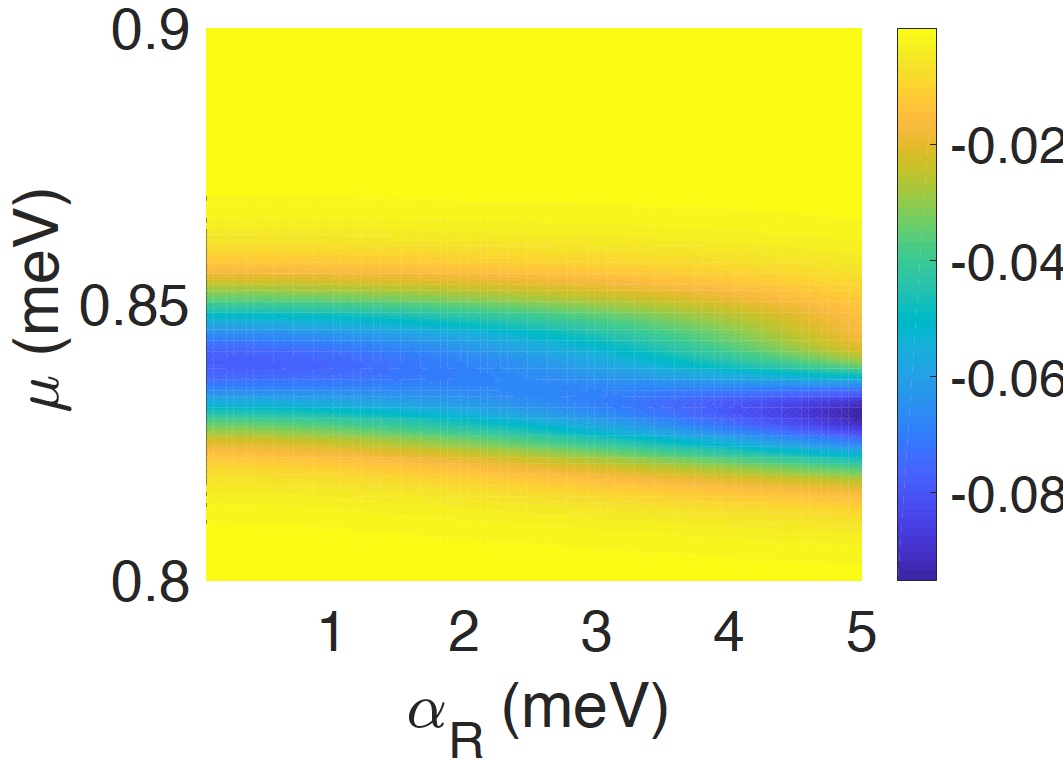}
	\caption{\textit{Top panel:} Conduction band spectrum of 2H stacked bilayer MoS$_2$ for $\alpha_R=0$ (left) and $\alpha_R=8meV$ (right). The band degeneracy is lifted by a finite Rashba soc. \textit{Bottom panel:} The conduction band Berry curvature (left) at the $\mathbf{K}$ point as a function of $\alpha_R$, which displays a giant peak for small finite $\alpha_R$. Anomalous valley Nernst coefficient (right) $\alpha_{xy}$ as a function of $\mu$ and $\alpha_R$. }
	\label{Fig_2H_bilayer_1}
\end{figure}
\begin{figure}
	\includegraphics[scale=.215]{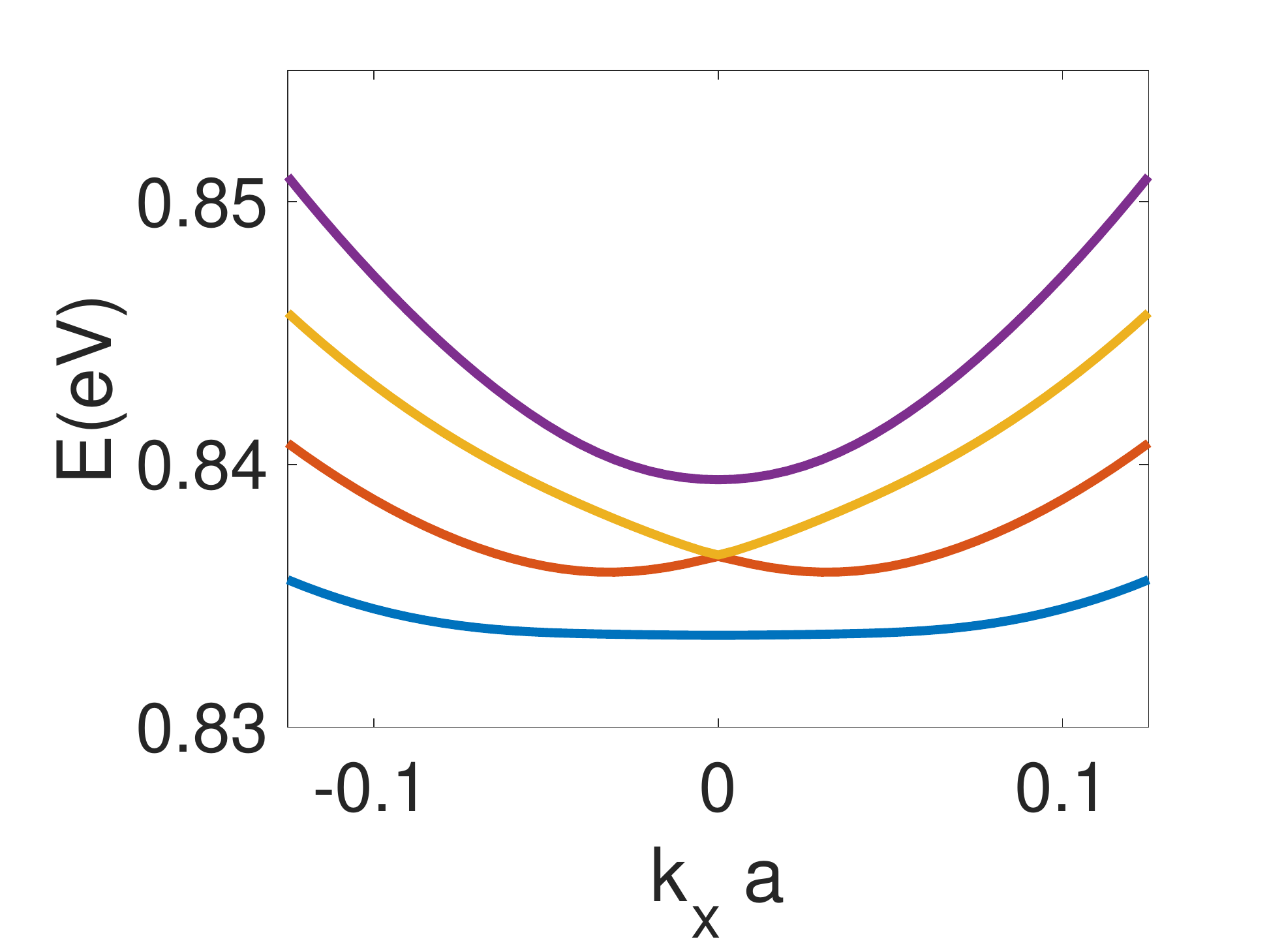}
	\includegraphics[scale=.215]{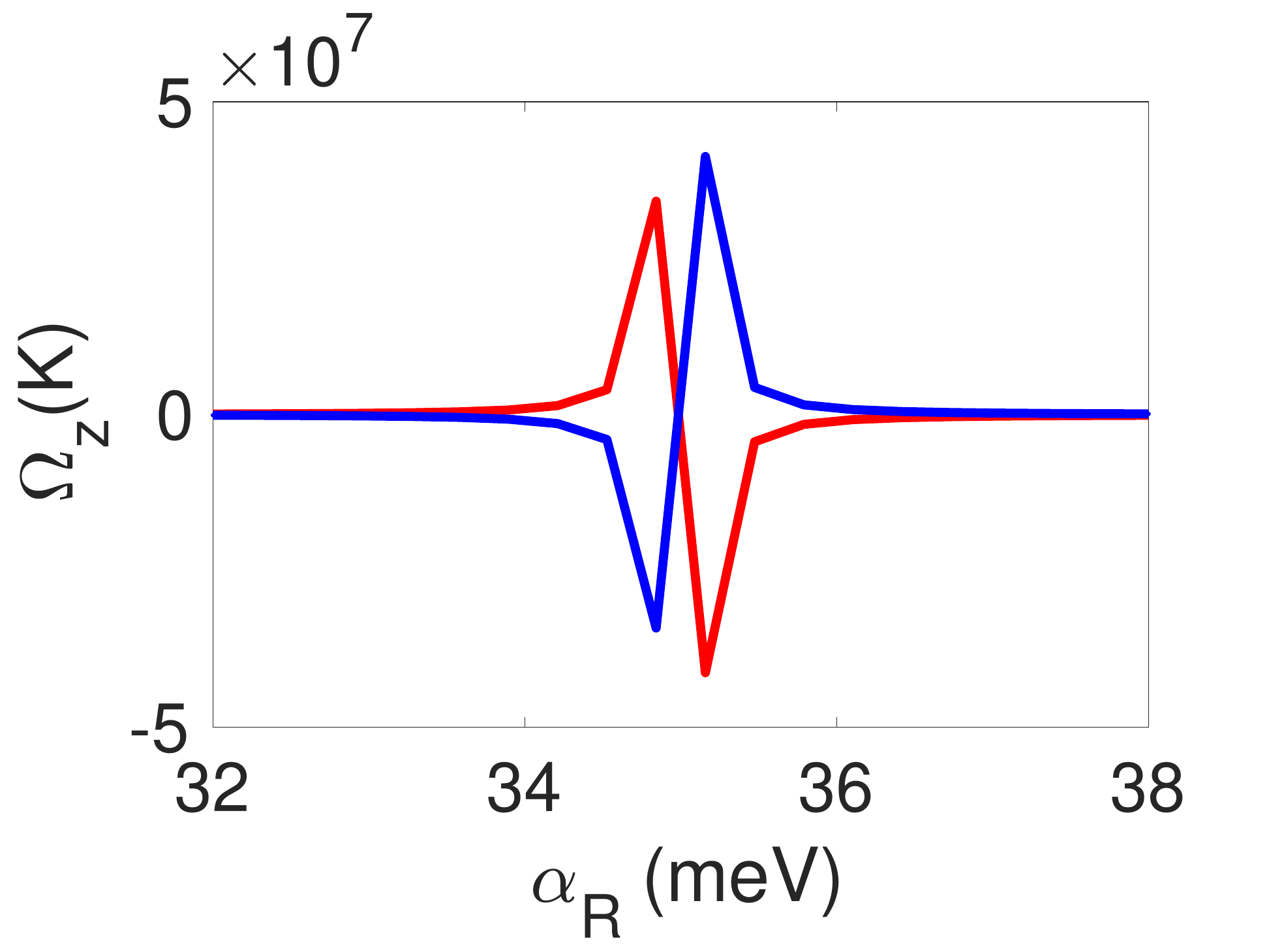}
	\includegraphics[scale=.21]{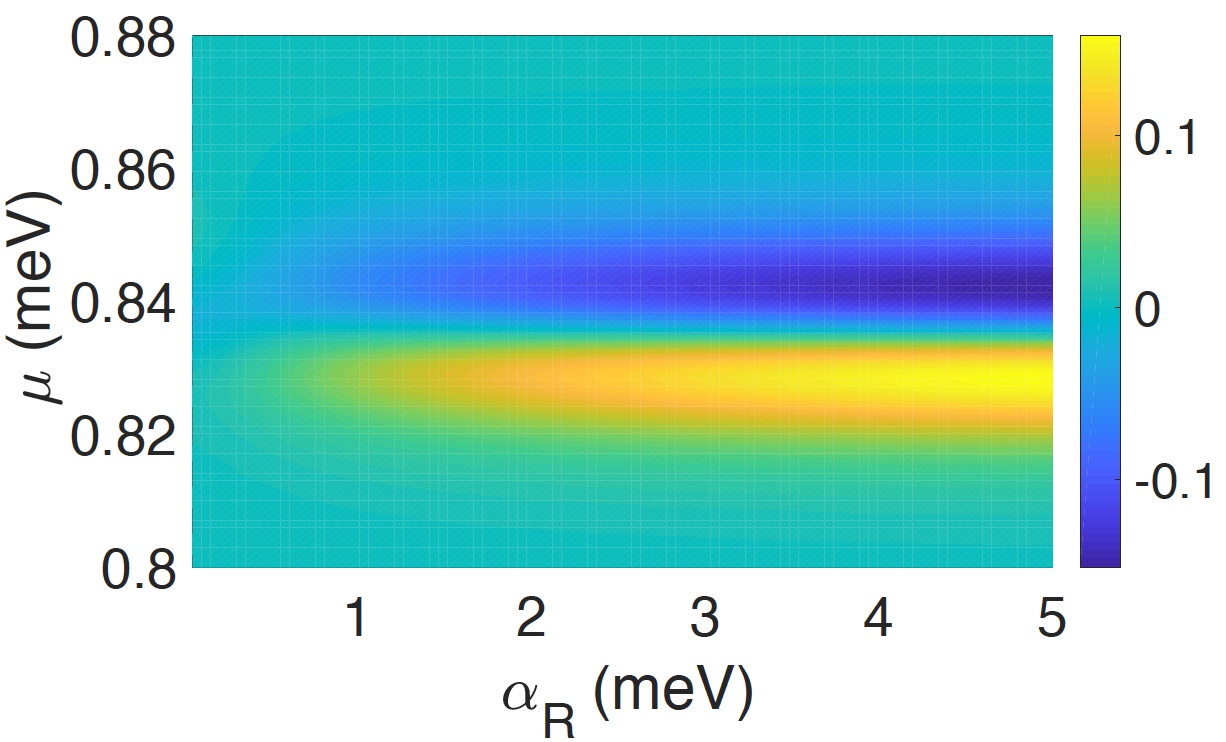}
	\caption{\textit{Top panel:} Conduction energy bands (left) and the Berry curvature in the units of $\AA^2$ (right) for 2H stacked bilayer TMD at large values of Rashba SOC. At certain value of $\alpha_R$, two of the bands cross each other leading to a peak in the Berry curvature along with a sign change. \textit{Bottom panel:} Total anomalous Nernst coefficient ($\alpha_{xy}$) of a 2H stacked bilayer MoS$_2$ for a magnetic field of $B=1T$.}
	\label{Fig_2H_bilayer_2}
\end{figure}

Let us now discuss the tunability of the Berry curvature in the presence of Rashba spin orbit coupling. Fig.~\ref{Fig_bilayer_1} shows the peak Berry curvature at the $\mathbf{K}$ point as a function of the Rashba spin orbit coupling parameter ($\alpha_R$) for four of the total eight bands of a 3R stacked bilayer MoS$_2$. The Berry curvature is easily tunable with Rashba spin orbit coupling and can be enhanced up to two orders of magnitude for typical $\alpha_R$ values. Similar to the case of a monolayer TMD, we can expect a large anomalous valley Nernst signal, which is easily tunable via Rashba spin-orbit coupling. Again, due to preservation of a TR symmetry, the net Nernst signal must vanish. A small perturbative magnetic field, can however break valley symmetry. The magnetic field acts on the valley and spin sectors i.e. $H^s_z = b_z s_z$, and $H^v_z = \sum\limits_i m^i b_z$, for magnetic field $b_z$ and valley orbital magnetic moments $m^i$. Although, the magnetic field has a small effect on the band-structures of these materials, its role in generating a valley asymmetry and thus a large overall Berry curvature is crucial. Fig.~\ref{Fig_bilayer_1} shows the total anomalous Nernst coefficient for a magnetic field of $5T$, which is tunable with the chemical potential and $\alpha_R$, showing sharp peaks at certain values of chemical potential. 

\subsection{2H stacked bilayer}
The low energy $\mathbf{k}\cdot \mathbf{p}$ Hamiltonian for 2H stacked bilayer MoS$_2$ is given by 
\begin{align}
H^{2H}=\left( \begin{array}{cccc}
\epsilon_{cb} & \gamma_3 k^+ & \gamma_{cc} k^- & 0 \\
\gamma_3 k^- &   \epsilon_{vb}   & 0 & t_{\perp} \\
\gamma_{cc} k^+ & 0 & \epsilon_{cb}   & \gamma_3 k^-\\
0 &t_{\perp}  & \gamma_3 k^+ &  \epsilon_{vb}
\end{array} \right)
\label{Eq_H_2H}
\end{align}
where $t_{\perp}$ is the tunneling between the holes of top and bottom layer. Unlike 3R stacked bilayer, the above Hamiltonian is inversion symmetric. Therefore the Berry curvature vanishes identically everywhere. The spin-orbit coupling effects can be accounted for by  $\Delta_{cb} s_z \tau_z \sigma_z$ and $\Delta_{vb} s_z \tau_z \sigma_z$ to the conduction and the valence bands respectively. However, even in the presence of SOC, the energy bands in 2H stacked bilayer remain degenerate (see Fig.~\ref{Fig_2H_bilayer_1}), leading to vanishing Berry curvature. A finite Berry curvature can be generated via: (i) a finite inter-layer potential $U_g$ which breaks symmetry between the top and the bottom layers, or (ii) gate-tunable Rashba spin-orbit coupling. The effect of inter-layer coupling has been explored in Ref.~\onlinecite{Kormanyos:2018}, and we will not discuss that here. Here we will explore the effect of Rashba spin-orbit coupling on the band-structure, Berry curvature and the Nernst effect in 2H stacked bilayer TMDs. 

The evolution of band spectrum due to Rashba SOC is shown in Fig.~\ref{Fig_2H_bilayer_1}. Even in the presence of intrinsic SOC (but absence of Rashba SOC), the 2H bilayer spectrum is spin-degenerate, with vanishing  Berry curvature near a valley point. A finite Rashba SOC lifts the spin degeneracy  at and away from the $\mathbf{K}$ point. We note that due to an infinitesimal Rashba SOC, the band degeneracy is lifted and a giant peak in the Berry curvature is generated. This behavior can be contrasted to the smooth Berry curvature enhancement in 3R stacked bilayer MoS$_2$ (see Fig.~\ref{Fig_bilayer_1}) when $\alpha_R$ is increased from zero. Consequently, an anomalous valley Nernst coefficient, tunable with Rashba SOC, is generated (also shown in Fig.~\ref{Fig_2H_bilayer_1}), displaying peaks when the chemical potential touches the energy bands. Another interesting behavior occurs for large values of Rashba SOC. When the Rashba SOC parameter is increased further, at a certain critical $\alpha_R$ the bands cross each other (see Fig.~\ref{Fig_2H_bilayer_2}). This band crossing also generates a peak in the Berry curvature along with a distinct sign change (Fig.~\ref{Fig_2H_bilayer_2}). Thus, even though the 2H stacked bilayer by itself does not display peculiar Berry curvature effects, a small finite Rashba coupling can lead to striking modulation of the Berry curvature. 
Now, we can apply a small magnetic field which breaks the valley symmetry, and thereby generates a large overall anomalous Nernst effect. Fig.~\ref{Fig_2H_bilayer_2} shows the total anomalous Nernst coefficient ($\alpha_{xy}$) of a 2H stacked bilayer MoS$_2$ for a magnetic field of $B=1T$, tunable with Rashba SOC.

\section{Conclusion}
 In this work we analyzed the Nernst response of monolayer and bilayer TMDs. The Nernst response typically has two electronic contributions - the conventional $B-$field dependent contribution, where the anomalous velocity to the electrons is provided by the Lorentz force, and secondly the the anomalous  contribution where the anomalous velocity is provided by the Berry curvature. 
 We first showed that the Nernst signal at a valley point in TMDs is purely dominated by the anomalous geometrical contribution and the conventional $B-$dependent Nernst signal is significantly weaker for typical material parameters. 
 Further, we showed that the interplay of gate controlled Rashba SOC and intrinsic Ising SOC in gated monolayer TMDs can produce a large valley Nernst signal of purely geometrical origin in monolayer TMDs, which is highly tunable in nature. 
The tunable anomalous valley Nernst signal in the presence of Rashba coupling can be enhanced by at least one order of magnitude. Although the total Nernst signal vanishes due to preservation of TR symmetry, a small magnetic coupling can lift the valley degeneracy, and an amplified net Nernst signal of almost two orders of magnitude larger can be achieved.
 
Additionally, we also analyzed the band structure, Berry curvature, and Nernst effect in bilayer TMDs, focusing on 3R and 2H stacked bilayer MoS$_2$ as a prototype. Including just the intrinsic SOC effects, the 3R stacked bilayer MoS$_2$ breaks inversion symmetry, unlike the 2H stacked bilayer which remains spin-degenerate. We showed that a finite Rashba SOC can enhance and tune the valley Berry curvature of a 3R stacked bilayer. On the other hand a small but finite Rashba SOC generates (and can also modulate) a giant Berry curvature in a 2H stacked bilayer due to lifting of spin-degeneracy. When the Rashba SOC is increased further, then at a certain critical $\alpha_R$ the bands cross each other and another peak in the Berry curvature is generated along with a distinct sign change. Consequently, we also showed that Rashba SOC can enhance and tune the anomalous Nernst effect in bilayer TMDs.
Our predictions can be directly tested in experiments, and may be employed for several promising tunable caloritronics applications.

%%%%%%%%%%%%%%%%%%%%%%%%%%%%%%%%%%%%%%%%%%%%%%%%%%%%%%%%%%%%%%%%%%%%%%%%

\end{document}